\documentclass[11pt]{article} 

\usepackage[a4paper,nohead,dvips,twoside]{geometry}
\usepackage{ucs}
\usepackage[utf8x]{inputenc}
\usepackage[english]{babel}
\usepackage{amsmath}
\usepackage{amssymb}
\usepackage[dvips]{graphicx}
\usepackage{psfrag}
\usepackage{tensor}
\usepackage{xspace}

\usepackage[ps2pdf, bookmarks=true,
            bookmarksopen=true,colorlinks=false]{hyperref}

\hypersetup{%
  pdftitle = {A spectral solver for evolution problems 
  with spatial S3-topology},
  pdfsubject = {},
  pdfauthor = {Florian Beyer},
  pdfkeywords = {}%
}

 \usepackage[thmmarks,amsmath,hyperref]{ntheorem}

 \theoremstyle{plain}
 \theorembodyfont{\normalfont}
 \theoremsymbol{\ensuremath{_\Box}}

 \theoremheaderfont{\sc}\theorembodyfont{\small\upshape}
 \theoremstyle{nonumberplain}
 \theoremseparator{:}
 \theoremsymbol{\rule{1ex}{1ex}}
 \theoremstyle{break}
 \theoremheaderfont{\normalfont\small}
 \theorembodyfont{\it\small\flushleft}
 \theoremsymbol{}
 
 \theorembodyfont{\it\tiny\flushleft}

\newcommand{\R}{\ensuremath{\mathbb R}\xspace}

\renewcommand{\Re}{\mathrm{Re}\,}
\renewcommand{\Im}{\mathrm{Im}\,}
\newcommand{\oplo}{\ensuremath{1+1}\xspace}
\newcommand{\tplo}{\ensuremath{2+1}\xspace}
\newcommand{\Sone}{\ensuremath{\mathbb S^1}\xspace}
\newcommand{\Stwo}{\ensuremath{\mathbb S^2}\xspace}
\newcommand{\SoXSt}{\ensuremath{\mathbb S^1\times\mathbb S^2}\xspace}
\newcommand{\Sth}{\ensuremath{\mathbb S^3}\xspace}

\newcommand{\T}{\ensuremath{\mathbb T^3}\xspace}
\newcommand{\U}{\ensuremath{\mathrm{U(1)}}\xspace}
\newcommand{\SU}{\ensuremath{\mathrm{SU(2)}}\xspace}

\newcommand{\normadapt}{\mathrm{Norm}^{(\textit{adapt})}}
\newcommand{\normdiff}{\mathrm{Norm}^{(\textit{diff})}}
\newcommand{\normbc}{\mathrm{Norm}^{(\textit{BC})}}
\newcommand{\normconstr}{\mathrm{Norm}^{(\textit{constr})}}
\newcommand{\normeinstein}{\mathrm{Norm}^{(\textit{einstein})}}

\newcommand{\Connection}[3]{\Gamma\indices{_{#1}^{#2}_{#3}}}

\newcommand{\tr}{\mathrm{tr}\,}
\newcommand{\scri}{\ensuremath{\mathcal J}\xspace}

\newcommand{\scrip}{\ensuremath{\mathcal J^+}\xspace}

\newcommand{\Eqref}[1]{Eq.~\eqref{#1}}
\newcommand{\Eqsref}[1]{Eqs.~\eqref{#1}}
\newcommand{\Sectionref}[1]{Section~\ref{#1}}

\newcommand{\Figref}[1]{Fig.~\ref{#1}}

\newcommand{\term}[1]{#1}

\graphicspath{{figures/}}

\begin{document}
\title{A spectral solver for evolution problems 
  with spatial \Sth-topology}

\author{Florian Beyer\\\textit{\small beyer@ann.jussieu.fr}}
  
\date{
{\small Laboratoire Jacques-Louis Lions\\
Universit\'e Pierre et Marie Curie (Paris 6)\\
4 Place Jussieu, 75252 Paris, France}}

\maketitle 

\begin{abstract}
We introduce a single patch collocation method in order to compute
solutions of initial value problems of partial differential equations
whose spatial domains are $3$-spheres. Besides the main ideas, we
discuss issues related to our implementation and analyze numerical
test applications. Our main interest lies in cosmological
solutions of Einstein's field equations. Motivated by this, we also
elaborate on problems of our approach for general tensorial evolution
equations when certain symmetries are assumed. We restrict to \U- and
Gowdy symmetry here.


\end{abstract}

\renewcommand{\figurename}{Fig.}
\bibliographystyle{hplain}

\section{Introduction}
Numerical studies of initial value problems of partial differential
equations on certain spatial domains have a long history both in basic
research and in applied science, see for instance \cite{boyd} and
references therein.  In particular, for applications in geometrical
classical theories in physics, as for instance general relativity,
Maxwell theory, but also Ricci flow (introduced e.g.\ in
\cite{Ricciflow}), there are interesting applications where the
spatial domain is a $3$-sphere. In general relativity, 
spatial \Sth-topology plays a particularly important role for the
standard model of cosmology based on the spatially homogeneous and
isotropic Friedmann-Robertson-Walker solutions (see for instance
\cite{hawking,Wainwright}). Beyond those simple and, at least for
certain matter fields, well-understood models with high symmetry,
there exist several outstanding open problems; of particular
outstanding interest and motivation for this work here are the strong
cosmic censorship and BKL conjecture in Gowdy vacuum solutions of
Einstein's field equations for spatial \Sth- or \SoXSt-topologies
\cite{Gowdy73,Isenberg89,Chrusciel90,garfinkle1999,beyer08:TaubNUT,Stahl02}.


It is not straight forward to deal with ``non-trivial'' spatial
topologies, such as \Sth, numerically. Recall the well-known problems
occurring at the coordinate axis in the case of standard cylindrical
coordinates $(\rho,\phi,z)$ in $\R^3$. These coordinates degenerate at
the ``axis'' given by $\rho=0$. In typical equations, this has the
consequence that derivatives with respect to the azimuthal angle
$\phi$ always come together with a factor $1/\rho$. If $f$ is a smooth
function on $\R^3$, then a term like $1/\rho\,\partial_\phi f$ is
well-behaved at the axis. However, when an equation with such terms is
solved numerically, the ``formal singularity'' $1/\rho$ can cause
numerical instabilities. Applications with axial symmetry have a great
history in all over science. Just to name examples of numerical
studies of axisymmetric problems in general relativity, we list
\cite{Garfinkle00,Choptuik03,Rinne05}. In this paper, we will not be
interested in axial symmetry, however, it is instructive to keep it in
mind for the following reason.  Let us assume that we cover a dense
subset of \Sth with one coordinate patch, for instance the Euler
coordinates introduced below.  Then it turns out that this leads,
loosely speaking, to two ``axes'' on \Sth, each of which with similar
properties as the axis for cylindrical coordinates on $\R^3$. Hence,
we call this the ``axis problem'' also in the case of spatial
\Sth-topology.

Alternatively to this approach with one singular coordinate patch on
the spatial domain, one can try a multipatch technique. The idea is to
cover the spatial domain with several local regular coordinate
patches. In general relativity, examples of implementations of the
multipatch technique are \cite{thornburg04,tiglio05}; of particular
interest for our work here are implementations based on spectral
methods in \cite{Pfeiffer02,Scheel06,duez2008}. In any case, the
multipatch technique does not seem advantageous for the applications
which we have in mind. First, its implementation is difficult, since
one must find a stable and efficient way of guaranteeing the necessary
communication between the local patches. Second, we are interested in
cosmological solutions with symmetries, and in order to take full
advantage of those, it is often a good idea to adapt the coordinates
to the symmetries, even though this can mean that one has to deal with
singular coordinate systems. We have decided to develop a single patch
code based on the collocation method\footnote{In this paper we will
  often speak sloppily of spectral or pseudospectral methods when we
  mean the collocation method.} for the spatial discretization. It is
general experience that such methods typically yield high accuracy
\cite{boyd}.  Furthermore, a spectral single patch approach seems very
natural from our geometric point of view, which we introduce in this
paper. For the time discretization, we use the method of lines with a
Runge Kutta integrator. Such a discretization technique for various
spatial domains has a long history in computational physics, see for
instance the references and examples in \cite{boyd}. An alternative
approach, which uses spectral discretization both in space and in
time, has been reported on in \cite{Hennig08}. However, to our
knowledge, the case of spatial \Sth-topology has not been studied yet.

Our aim is to find numerical solutions of systems of first order
quasi-linear evolution PDEs, written schematically as
\begin{equation}
  \label{eq:abstractevolution}
  \partial_t f(t,x)
  +\sum_i(A^i(f,t,x)\cdot\nabla_i)f(t,x)+B(t,x,f)=0.
\end{equation}
Here, the unknown $f$ is a vector, the terms $B$ are vector valued,
and $A^i$ and $\nabla_i$ represent a collection of matrices and
spatial derivative operators, as explained in more detail later. The
spatial domain, represented by the abstract coordinates $x$, is
\Sth. We assume in the following, without further notice, that the
initial value problem for such a system is well-posed, i.e.\ that for
any choice of initial data in a given regularity class, there exists a
unique solution $f$ locally in time, which depends continuously on the
initial data in a well-defined manner. In our applications, ignoring
the ``axis problem'' above for the moment, the operators $\nabla_i$
can be thought of as the spatial partial derivatives, the quantities
$A^i$ as symmetric matrices, and all coefficients depend smoothly on
their arguments. In this case, the system is symmetric hyperbolic, and
well-posedness follows \cite{john82,Majda84}. In particular, when the
initial data is smooth, then the solution is smooth until it breaks
down. We remark, that in principle our approach applies to other
forms of hyperbolicity for \Eqref{eq:abstractevolution}, or even
parabolic systems. 

We will often use the following equivalent geometric language. Namely,
we say that we look for solutions of \Eqref{eq:abstractevolution} on
$\R\times\Sth$, where time $t$ is interpreted as the canonical
coordinate on the factor $\R$ of the manifold $\R\times\Sth$, and the
$t=const$-hypersurfaces have \Sth-topology.  In this paper, we will
often be concerned with the case when $f$ in
\Eqref{eq:abstractevolution} represents components of smooth tensor
fields. Then we call \Eqref{eq:abstractevolution} ``tensorial
equation''.

This paper is organized as
follows: \Sectionref{sec:geomsetupnumapproach} is devoted to the
description and discussion of our numerical technique.  We show our
geometric point of view which leads, in a natural way, to a single
patch collocation discretization in space. We demonstrate that it
allows a straight forward treatment of the ``axis problem'' mentioned
above. Since we are interested, in particular, in tensorial equations,
we discuss several related issues
in \Sectionref{sec:tensorialequ}. Namely, in order to express the
tensor fields as collections of smooth functions we have decided to work
with smooth global frames on \Sth. The formulation of tensorial
equations in terms of smooth global frames on \Sth leads to certain
problems in the presence of symmetries. Two particular classes of
symmetries of our interest are discussed. For these sections, but
indeed at many places of this paper, some background in differential
geometry would be helpful, which can be obtained for instance from
\cite{hawking,oneil}. In \Sectionref{sec:implementation}, we elaborate
on our numerical infrastructure in general, and we discuss certain
issues related to numerical stability which are present for spatial
\Sth-topology. In \Sectionref{sec:testsresults}, we proceed as
follows. First, we introduce the mathematical and physical background
of our test application. Then we test the code in that setting and
elaborate on numerical errors, stability and performance. Finally,
in \Sectionref{sec:discussion}, we summarize and conclude. We comment
briefly on possible problems when some of our special assumptions are
dropped, point to open issues and list aspects for future work.


\section{Geometric ideas and numerical implementation}
\label{sec:geomsetupnumapproach}

\subsection{Euler coordinates and the Euler map}
\label{sec:geometry}
In the following, we consider the $3$-sphere \Sth as the submanifold
of $\R^4$ given by
\begin{equation}
  \label{eq:S3}
  \Sth:=\left\{(x_1,x_2,x_3,x_4)\in\R^4\,
    \left|\,\sum_{i=1}^4x_i^2=1\right.\right\}.
\end{equation}
\begin{subequations}%
  \label{eq:euler_param}%
  The Euler coordinates of $\Sth$, also appearing as Euler angle
  parametrization, etc.\ in the literature, is the coordinate patch
  given by
  \begin{gather}
    \begin{split}
      x_1&=\cos\frac{\chi}2\cos\lambda_1, \quad 
      x_2=\cos\frac{\chi}2\sin\lambda_1,\\
      x_3&=\sin\frac{\chi}2\cos\lambda_2,\quad
      x_4=\sin\frac{\chi}2\sin\lambda_2,
    \end{split}
  \end{gather}
  in terms of the coordinate functions $\chi\in[0,\pi]$,
  $\lambda_1,\lambda_2\in\left[0,2\pi\right[$.  We will rather use the
  coordinates $\{\chi,\rho_1,\rho_2\}$ determined by
  \begin{equation}
    \lambda_1=(\rho_1+\rho_2)/2,\quad\lambda_2=(\rho_1-\rho_2)/2.
  \end{equation}
\end{subequations}
The Euler coordinates smoothly cover the dense subset of \Sth given,
when the points $\chi=0,\pi$ are taken away. We expect that other
choices of coordinates with similar properties are also appropriate.
Although the motivation for choosing the Euler coordinates stems from
Gowdy symmetry, as becomes clearer later, they are robust enough for
more general cases.

Certainly, the relations in \Eqref{eq:euler_param} are well-defined
for all $\chi,\rho_1,\rho_2\in\R$. This is also true when we consider
$\chi,\rho_1,\rho_2\in (\R\text{ mod } 4\pi)$.  Geometrically, the
Euler coordinates given by \Eqref{eq:euler_param} can thus be
interpreted as a smooth map from the $3$-torus
\[\T:=(\R\text{ mod } 4\pi)\times(\R\text{ mod } 4\pi)
\times(\R\text{ mod } 4\pi)\] to \Sth, which we call the Euler map
\begin{equation}
  \label{eq:Euleranglemap}
  \begin{split}
    &\Phi: \T\rightarrow\Sth,\\
    (\chi,\rho_1,\rho_2)\mapsto
    &\left(\cos\frac{\chi}2\cos\frac{\rho_1+\rho_2}2,\quad
    \cos\frac{\chi}2\sin\frac{\rho_1+\rho_2}2,\right.\\
    &\left.\sin\frac{\chi}2\cos\frac{\rho_1-\rho_2}2,\quad
    \sin\frac{\chi}2\sin\frac{\rho_1-\rho_2}2\right)\in\Sth.
  \end{split}
\end{equation}
The Euler map $\Phi$ is even a diffeomorphism when we restrict it to
e.g.\ $\chi\in]0,\pi[$ and restrict the image correspondingly. But at
the points $\chi=0,\pi$, the inverse is not well-defined.  Note, that
in the whole paper we will often make the canonical identification of
the isomorphic groups $\U$, $\Sone$, $(\R\text{ mod } 2\pi)$ and
$(\R\text{ mod } 4\pi)$, and henceforth not distinguish between
them. However at this point, our definition of the map requires to
stand on $4\pi$-periodicity at least for the coordinate $\chi$, but we
come to the standard $2\pi$-periodicity in a moment.

Let $f$ be a smooth function on \Sth. In the following we consider
$\tilde f:=f\circ\Phi$, where $\Phi$ is the Euler map defined in
\Eqref{eq:Euleranglemap}. Hence $\tilde f$ is a smooth function on
\T. For simplicity, we write $f$ instead of $\tilde f$, and often make
no difference between the ``original function $f$ on $\Sth$'' and the
``corresponding function $\tilde f$ on $\T$''. If necessary, in order
to avoid confusions, we sometimes say that ``$f$ is a smooth function
on \T originating in a smooth function on \Sth''.

Motivated by this simple geometric relation given by the Euler map,
our approach is the following one; the details are worked out in the
subsequent sections. Our aim is to solve partial differential
equations with a spatial domain \Sth.  Let us suppose that all
coefficient functions and unknown functions, which we want to solve
for, in the equation are smooth functions on \Sth. Furthermore,
suppose that all derivative operators stem from smooth globally
defined vector fields on \Sth.  Since $\Phi$ becomes a diffeomorphism,
when we restrict it to a dense subset of \Sth, all these functions and
vector fields correspond in a unique manner to smooth functions and
vector fields on the corresponding subset of \T. Hence, we can solve
the equation as if it its spatial domain were \T. However, because
$\Phi$ is not a diffeomorphism globally, we have introduced formal
singularities to the equations, analogous to the singularities at the
axis of cylindrical coordinates on $\R^3$ discussed above.  One main
conclusion in the following sections is that the hypothesis, that all
these quantities on \T originate in smooth quantities on \Sth, allow
to regularize the formally singular behavior at those points.  By all
this, we will successfully make ``\Sth periodic in all three spatial
directions'', and can henceforth use spectral methods based on the
standard Fourier basis for the spatial discretization of the
equations.

In all of what follows, we will assume, for simplicity, that all
functions involved do not depend on the coordinate $\rho_2$. We call
such functions \U-symmetric, and later we interpret this symmetry
geometrically. The generalization of the following ideas is, however,
straight forward. This assumption has the following nice property. Let
$f$ be a smooth \U-symmetric function on \Sth. Then $f\circ\Phi$ is
$2\pi$-periodic (instead of $4\pi$-periodic) in $\chi$ and
$\rho_1$. This is so, because we can use the symmetry in $\rho_2$ to
switch the signs of all terms in \Eqref{eq:Euleranglemap} consistently
as needed. Hence, in the following, we only need to deal with the
standard $2\pi$-periodicity.

Note, that analogous spectral approaches for equations with spatial
domains diffeomorphic to \Stwo have been implemented before, see for
instance \cite{Bartnik1999,boyd} and references therein.

\subsection{Analysis of Fourier series of given smooth functions on
  \texorpdfstring{\Sth}{S3}}
\label{sec:analysisfourier}

Let $f$ be a given smooth \U-symmetric function on \Sth. The
corresponding function $f$ on \T is smooth, and hence has a
representation in terms of a Fourier series. We come back to the
question of convergence of such a series later.  At this stage, we
want to assume that this Fourier series is finite, and, without loss
of generality, with $N$ summands both in $\chi$ and $\rho_1$.  Since
$f$ is \U-symmetric function, it is $2\pi$-periodic in both $\chi$ and
$\rho_1$, as argued before. Let $f$ be real valued. Then it must be of
the form
\begin{equation}
  \label{eq:Fourier_smoothapprox2}   
  f(\chi,\rho_1)=F_0(\chi)
  +2\,\text{Re}\sum_{p=1}^{N}F_p(\chi)e^{ip\rho_1}.
\end{equation}
\begin{subequations}%
  \label{eq:Fourier_smoothapprox3}%
  Here, $F_0(\chi)$ is a real valued function, and $F_p(\chi)$ is
  allowed to be complex for all $p\ge 1$.  In \cite{beyer:PhD}, we
  demonstrated that the properties of the Euler map under \U-symmetry
  implies
  \begin{equation}
    \label{eq:FNpFourier01}
    F_p(\chi)=
    \begin{cases}
      {\displaystyle 2 \sum_{n=1}^{N} f_{n,p}\cos n\chi+f_{0,p}} &
      \text{for $p\ge0$ even},\\
      {\displaystyle -2 i \sum_{n=1}^{N} f_{n,p}\sin n\chi} &
      \text{for $p>0$ odd},
    \end{cases}
  \end{equation}
  for some, in general, complex coefficients $f_{n,p}$; only the
  coefficients for $p=0$ must be real. Some of the factors in these
  expressions are chosen for later convenience. Note, that in
  \cite{beyer:PhD}, the coordinates $(\chi,\rho_1,\rho_2)$ are defined
  slightly differently. For example, the function $\chi$ here must be
  substituted by $2\chi$ to compare to the expressions in
  \cite{beyer:PhD}. In any case, we further find that for even $p>0$,
  there are the ``compatibility conditions''
  \begin{equation}
    \label{eq:approximation_comp_cond}
    f_{0,p}+2\sum_{n=1}^{N} f_{2n,p}=0,\quad 
    \sum_{n=1}^{N} f_{2n-1,p}=0.
  \end{equation}
\end{subequations}
These originate in the fact that the functions $F_p(\chi)$ must vanish
at the degenerated places $\chi=0,\pi$ for all $p>0$. For odd $p$,
this is implied by the expression \Eqref{eq:FNpFourier01}
automatically, and hence there are no compatibility conditions.  The
corresponding expressions are more complicated, but analogous, when we
give up \U-symmetry.

Now, we want to study what happens to the Fourier series of $f$, when
it is differentiated along a smooth tangent vector field on \Sth. By
this, we mean that the abstract derivative operator $\nabla_i$ in
\Eqref{eq:abstractevolution} is of the form $\nabla_i
f=V^\alpha\partial_{x^\alpha} f$ for a smooth tangent vector field
$V=V^\alpha\partial_{x^\alpha}$ on \Sth. Here, our convention is that
$x^\alpha$ represents three abstract spatial coordinates, which
neither need be the Cartesian coordinates, nor the Euler coordinates
used before. However, we will restrict to Euler coordinates in the
following. Furthermore, we assume Einstein's summation convention. The
coefficients $V^\alpha$ are just functions on \Sth which are not
necessarily smooth when we consider the Euler coordinates.  Of
particular importance will be the following tangent vector fields,
whose origin we explain later and which, with respect to the Euler
coordinates, take the form
\begin{subequations}%
  \label{eq:coordinate_repr_standard_frameY}%
  \begin{align}
    Y_1&=2\sin \rho_1\,\partial_\chi
    +2\cos \rho_1\,
    \left(\cot\chi\,\partial_{\rho_1}-\csc\chi\,\partial_{\rho_2}\right),\\
    Y_2&=2\cos \rho_1\,\partial_\chi
    -2\sin \rho_1\,
    \left(\cot\chi\,\partial_{\rho_1}-\csc\chi\,\partial_{\rho_2}\right),\\
    Y_3&=2\partial_{\rho_1}.        
  \end{align}
\end{subequations}
The factors $2$ are chosen for consistency with our discussion in
\cite{beyer:PhD}.  Now, as we explain later, any smooth vector field
$V$ on \Sth can be written as a linear combination $V=V^a Y_a$ with
$V^1$, $V^2$, $V^3$ smooth functions on \Sth. Hence, under the
assumption that all differential operators in the equations stem from
smooth vector fields on \Sth, it follows for \U-symmetry, that all
``formally singular'' differential operators in our equations must be
of the form $-F(\chi,\rho_1)\cot\chi\partial_{\rho_1}$, with $F$ some
smooth function on \Sth. Without the assumption of $\U$-symmetry,
there can additionally be singular operators of the form
$(F(\chi,\rho_1,\rho_2)/\sin\chi)\partial_{\rho_2}$. This shows that
the formally singular terms here are of the same type as in the ``axis
problem''. The differences to the case of cylindrical coordinates on
$\R^3$ are twofold. First, in the case of \Sth, we have two such
``axes'' simultaneously at $\chi=0$ and $\pi$. Second, the axis itself
is not topologically a line here, but a closed circle. In the
following, we restrict our attention to the operator relevant to
\U-symmetry $-\cot\chi\partial_{\rho_1}$.

So, let $f$ be as before, and $g:=\partial_{\rho_1}f$. Since $g$ is
again a smooth $\U$-symmetric function on $\Sth$ with finite Fourier
series, the analogue of \Eqsref{eq:Fourier_smoothapprox2} and
\eqref{eq:Fourier_smoothapprox3} holds with $F_p(\chi)$ substituted
schematically by the function $G_p(\chi)$.  Then, for
$\chi\not=0,\pi$, we get,
\begin{subequations}%
  \label{eq:fouriersingular}%
  \begin{equation}
    \begin{split}
      &-\cot\chi\, G_p(\chi)\\
      &=
      \begin{cases} 
        {\displaystyle
          \begin{split}
            2\Biggl[
            \mathfrak{c}_{2,p}\sin\chi
            +\sum_{k=1}^{N}&\Bigl\{
            (\mathfrak{b}_{k,p}+\mathfrak{b}_{k+1,p})\sin 2k\chi\\
            &+(\mathfrak{c}_{k+1,p}+\mathfrak{c}_{k+2,p})\sin (2k+1)\chi
            \Bigr\}
            \Biggr]
          \end{split}}
        &\text{for $p>0$ even},\\
        {\displaystyle 
          \begin{split}
            2i\Biggl[
            \mathfrak{b}_{1,p}+
            \sum_{r=1}^{N}\Bigl\{&
            (\mathfrak{c}_{r,p}+\mathfrak{c}_{r+1,p})\cos(2r-1)\chi\\
            &+(\mathfrak{b}_{r,p}+\mathfrak{b}_{r+1,p})\cos 2r\chi
            \Bigr\}
            \Biggr]
          \end{split}
        } &\text{for $p>0$ odd}.
      \end{cases}
    \end{split}
  \end{equation}
  Here, we define
  \begin{equation}
    \label{eq:coeffSingTerms}
    \mathfrak{b}_{r,p}^{N}:=\sum_{n=r}^{N}g_{2n,p},
    \quad
    \mathfrak{c}_{r,p}:=\sum_{n=r}^{N}g_{2n-1,p}\quad\text{for }
    r\ge 1.
  \end{equation}
\end{subequations}
The computations leading to this result are described in
\cite{beyer:PhD}.  This result means that, as soon as the Fourier
coefficients of $\partial_{\rho_1}f$ are known, the Fourier
coefficients of the complete ``formally singular term''
$-\cot\chi\,\partial_{\rho_1}f$ can be computed.

Now, let us consider the general case of a given smooth \U-symmetric
function $f$ on $\Sth$. The associated function on \T has an infinite
Fourier representation in a general, with rapidly decreasing
coefficients $f_{n,p}$. This is a standard result from Fourier
analysis, which can be found in \cite{sugiura,canuto88}. This last
property means that the modules of the Fourier coefficients is bounded
by a uniform constant times any negative integer power of the two
summation indices $n$ and $p$. This property is often referred to as
``exponential convergence''. It is a general fact under our conditions
that the Fourier series converges pointwise absolutely and even
uniformly.  We find straight forwardly that the Fourier series of $f$
must be of the form given by \Eqsref{eq:Fourier_smoothapprox3},
setting $N\rightarrow\infty$. The infinite series of the compatibility
condition \Eqref{eq:approximation_comp_cond} converges because the
coefficients are rapidly decreasing.

Now consider the inverse question. Let a function be given on $\T$, of
the form of \Eqsref{eq:Fourier_smoothapprox2} and
\eqref{eq:Fourier_smoothapprox3}, for $N=\infty$ with rapidly
decreasing coefficients $f_{n,p}$. The standard theory implies that
the series converges pointwise absolutely and uniformly to a smooth
function $f$ on \T. However, does $f$ originate in a smooth function
on \Sth? In general the answer is no, because the compatibility
conditions \Eqref{eq:approximation_comp_cond} are just necessary, but
not sufficient for smoothness. Indeed, it is sufficient that for any
$p$, the function $F_p(\chi)$ is a smooth $2\pi$-periodic function on
\R, which has a zero of order $p$ at $\chi=0$ and $\pi$. In
particular, each function $F_p(\chi) e^{ip\rho_1}$ then originates in
a smooth function on \Sth. The argument for proving that this implies
that $f$ is a smooth function on \Sth, uses the theorem about Fourier
series on \Sth in \cite{sugiura}. Namely one can show that under
these assumptions, $f$ as a function on $\Sth$ can be represented as
an infinite series of spin-weighted spherical harmonics with again
rapidly decreasing coefficients.

Consider the derivative $g=\partial_{\rho_1}f$. In particular, the
formula for $-\cot\chi\,g$ in \Eqsref{eq:fouriersingular} also holds
in the limit $N\rightarrow\infty$, and the series expressions there
converge at least pointwise at all $\chi\not=0,\pi$. The function
$-\cot\chi\,\partial_{\rho_1}f$ is a smooth function on $\T$ because
each $F_p(\chi)$ in \Eqref{eq:Fourier_smoothapprox2} is a smooth
$2\pi$-periodic function in $\chi$ with appropriate zeros at the
``singular places'' $\chi=0,\pi$. This is a nice result because it
shows that \Eqref{eq:fouriersingular} is meaningful at the singular
locations, and hence allows to evaluate the formally singular term
$-\cot\chi\,\partial_{\rho_1}f$ explicitly there, even in the limit
$N\rightarrow\infty$. However, we remark that
$-\cot\chi\,\partial_{\rho_1}f$ does not originate in a smooth
function on \Sth, because \Eqsref{eq:fouriersingular} is not
consistent with \Eqref{eq:Fourier_smoothapprox3}. This is not a
problem because the formally singular operator is only a part of a
differential operator defined by a smooth vector field on
\Sth. Indeed, the result, when this ``full'' differential operator is
applied to a smooth function, yields a smooth function on \Sth.

\subsection{Symmetry and related issues for tensorial equations}
\label{sec:tensorialequ}

Before we discuss, how these results can be applied in practice, let
us first consider some consequences for tensorial equations.

\subsubsection{Smooth frames on \texorpdfstring{\Sth}{S3}}
Recall, that one of our main assumptions is that, at any given
instance of time, all unknowns and coefficients in the equations are
smooth functions, and that all differential operators are determined
by smooth globally defined vector fields on \Sth.  However, in order
to turn tensorial equations into partial differential equations for
smooth scalar functions, we need to introduce smooth frames on \Sth.
We would like to mention that an alternative way of treating tensorial
equations on \Sth in the case of Gowdy symmetry, see below, can be
found in \cite{garfinkle1999} for the case of spatial \SoXSt-topology.

Let us recall some well-known facts. Let $\Sth$ be given as in
\Eqref{eq:S3}.  Assume the standard matrix representation of the Lie
group \SU \cite{sugiura}.  The map
\begin{equation*}
  \Psi:\Sth\rightarrow\SU,\quad (x_1,x_2,x_3,x_4)\mapsto 
  \begin{pmatrix}
    x_1+i x_2 & -x_3+i x_4\\
    x_3+i x_4 & x_1-i x_2
  \end{pmatrix}
\end{equation*}
is a diffeomorphism, which can be used to transport the group
structure of $\SU$ to $\Sth$. Hence, both $\SU$ and $\Sth$ can be
considered as identical Lie groups via the map $\Psi$.  Thus, from the
standard \SU group multiplication, we can define left and right
translation maps,
\[L, R: \Sth\times\Sth\rightarrow\Sth,\quad (u,v)\mapsto L_u(v):=uv,
\quad (u,v)\mapsto R_u(v):=vu,\] so that $L_u$ and $R_u$ are
diffeomorphisms $\Sth\rightarrow\Sth$ for each point $u\in\Sth$.
Those maps can be employed to construct smooth global frames. First
one chooses a basis of the tangent space at the unit element $e$ of
the group. We choose the Pauli matrices with non-standard
normalization
\begin{equation*}
  \tilde Y_1=\begin{pmatrix}0&i\\i&0\end{pmatrix},\quad
  \tilde Y_2=\begin{pmatrix}0&-1\\1&0\end{pmatrix},\quad
  \tilde Y_3=\begin{pmatrix}i&0\\0&-i\end{pmatrix},
\end{equation*}
considered as elements of $T_e(\SU)$.  Then, one uses the push forward of
$L_u$ or $R_u$ to transport this basis smoothly to any other point
$u\in\SU$
\begin{equation*}
  (Y_a)_u:=(L_u)_*(\tilde Y_a),\quad (Z_a)_u:=(R_u)_*(\tilde Y_a).
\end{equation*}
Clearly, $\{Y_a\}$ is \SU-left invariant while $\{Z_a\}$ is \SU-right
invariant and both are smooth global frame fields on $\Sth$. It is
straight forward to check that they satisfy
\begin{equation}
  \label{eq:commutator_rel_Y}
  {}[Y_a,Y_b]=2\sum_{c=1}^3\epsilon\indices{_a_b_c}Y_c,
  \quad {}[Z_a,Z_b]=2\sum_{c=1}^3\epsilon\indices{_a_b_c}Z_c,
  \quad [Y_a,Z_b]=0,
\end{equation}
where $\epsilon\indices{_a_b_c}$ is the totally antisymmetric symbol
with $\epsilon\indices{_1_2_3}=1$. Here, the brackets denote the Lie
bracket. For $\{Y_a\}$, we have already written the explicit
expressions with respect to the Euler coordinates in
\Eqsref{eq:coordinate_repr_standard_frameY}. For $\{Z_a\}$, we have
\begin{subequations}%
  \label{eq:coordinate_repr_standard_frameZ}%
  \begin{align}
    Z_1&=-2\sin \rho_2\,\partial_\chi
    -2\cos \rho_2\,
    \left(\cot\chi\,\partial_{\rho_1}-\csc\chi\,\partial_{\rho_2}\right),\\
    Z_2&=2\cos \rho_2\,\partial_\chi
    -2\sin \rho_2\,
    \left(\cot\chi\,\partial_{\rho_1}-\csc\chi\,\partial_{\rho_2}\right),\\
    Z_3&=2\partial_{\rho_2}.
  \end{align}
\end{subequations}

On $\R\times\Sth$ with a time function $t$, we assume that each
$t=const$-hypersurface is diffeomorphic to \Sth with Euler coordinates
$\{\chi,\rho_1,\rho_2\}$. Hence, on each of these surfaces, the
expressions \Eqsref{eq:coordinate_repr_standard_frameY} and
\eqref{eq:coordinate_repr_standard_frameZ} define the fields $\{Y_a\}$
and $\{Z_a\}$. Geometrically, we thus have for all $a=1,2,3$
\[[\partial_t,Y_a]=[\partial_t,Z_a]=0.\]

Now we write an arbitrary globally defined smooth frame $\{e_i\}$ on
$\R\times\Sth$ as follows. By $\{e_i\}$ we mean the collection of $4$
frame fields $\{e_0,e_1,e_2,e_3\}$.  We set
\begin{subequations}
  \label{eq:frame}
  \begin{equation}
    e_0=\partial_t,
  \end{equation}
  and write,
  \begin{equation}
    \label{eq:orth_frame_Ya}
    e_a=e\indices{_a^b}Y_b,
  \end{equation}
\end{subequations}
where $(e\indices{_a^b})$ is a smooth $3\times 3$-matrix valued
function with non-vanishing determinant on \Sth. Our conventions for
frames is that the index $0$ always corresponds to the ``time frame
vector'' $e_0$ and $a,b,\ldots=1,2,3$ correspond to the ``spatial
frame vectors''. When we write indices $i,j,\ldots=0,\ldots,3$, we
mean both the time and spatial frame vectors.  For a tensor field $S$,
say for example a covariant $2$-tensor, we write $S_{ij}:=S(e_i,e_j)$,
$S_{ab}:=S(e_a,e_b)$ etc. We stress, that it is important to
understand, that writing $e_i$, does not mean the $i$th component of a
vector $e$, but rather the $i$th vector field of the frame $\{e_i\}$.

\subsubsection{Symmetry reductions of tensorial equations}
Let $S$ be an arbitrary smooth tensor field on $\R\times\Sth$.  We say
that $S$ is $\xi$-invariant, provided $\mathcal L_\xi S=0$
everywhere. Here $\mathcal L_\xi$ denotes the Lie derivative along
$\xi$. The coefficients $S_{ij}$ of $S$, with respect to an arbitrary
frame $\{e_i\}$, are constant along $\xi$, if, and in general only if,
$[\xi,e_i]=0$. Now, suppose that the functions $S_{ij}$ of such a
tensor are among the unknowns of the system of partial differential
equations which we would like to solve. Often, we would like to
exploit the symmetry of the unknowns and reduce the equations to some
simpler form. If the unknown functions are constant along $\xi$, such
a reduction can be done directly, if $\xi$ has the meaning of a
spatial coordinate vector field. Hence, in the following, we will be
interested in frames such that $[\xi,e_i]=0$; in this case, we say
that the frame is $\xi$-invariant.

Let us consider special cases of interest for us. We have already
introduced the notion of \U-symmetry for functions before. For general
smooth tensor fields $S$ on $\R\times\Sth$, we define it by the
requirement that $S$ is $Z_3$-invariant. One can define $\U$-symmetry
more geometrically, but for the purpose of this paper, our definition
is sufficient. The integral curves of $Z_3=\partial_{\rho_2}$ are
circles. Hence the symmetry group is isomorphic to \U, which motivates
the name. Now, it is straight forward to construct $Z_3$-invariant
frames $\{e_i\}$. Namely, for the ansatz \Eqsref{eq:frame}, this
requirement is equivalent to
\[\partial_{\rho_2}(e\indices{_a^b})=0,\]
due to \Eqsref{eq:commutator_rel_Y}.  The consequence is, that both
the frame matrix $(e\indices{_a^b})$ and the frame components of all
tensor fields, which are $Z_3$-invariant, are constant along $\rho_2$,
and hence are \U-symmetric functions. Provided, our equations are
formulated with respect to functions with this property only, the
spatial domain reduces to that of the coordinates
$\{\chi,\rho_1\}$. We refer to this as the \tplo-reduced equations.

Another symmetry assumption of interest is Gowdy symmetry with two
spatial symmetry vector fields. We say, that a tensor field is Gowdy
symmetric, if it is \U-symmetric and additionally $Y_3$-invariant.
Again, this definition can be made more geometric. In any case, let us
suppose that a frame field $\{e_i\}$, obeying \Eqsref{eq:frame} and
$[Z_3,e_i]=0$, is given as before. Now, it turns out, as argued in
\cite{beyer:PhD}, that the assumption that $\{e_i\}$ is a smooth
globally defined $Z_3$-invariant frame on \Sth, does not allow that it
is $Y_3$-invariant in addition. Hence, although we can find a frame
such that the frame components of arbitrary Gowdy symmetric tensor
fields are constant along $\rho_2$, it is not possible to achieve that
in addition they are constant along $\rho_1$; recall
$Y_3=\partial_{\rho_1}$. Thus, even if we assumed Gowdy symmetry, the
spatial domain of our equations would not reduce directly to that of
the coordinate $\chi$, i.e.\ not to \oplo dimensions. This difficulty
is a consequence of our assumption that the frame is smooth globally
on the manifold; if we allowed the frame to become singular at some
places, then the situation would be different. However, we would get
other problems due to the additional singularities.

Nevertheless, even under these assumptions, it is possible to perform
the following ``indirect'' reduction of the equations to \oplo
dimensions in the case of Gowdy symmetry. For the frame components of
any smooth tensor field $S$, for example in the case of a covariant
$2$-tensor, which is $Y_3$-invariant, we find
\begin{equation}
  \label{eq:Y3ofsymmetricTensor}
  Y_3(S_{ij})=S_{i'j}T\indices{_i^{i'}}+S_{ij'}T\indices{_j^{j'}}
\end{equation}
with 
\[T\indices{_i^{i'}}e_{i'}:=[Y_3,e_i].\]
Under our assumptions for the frame \Eqsref{eq:frame}, we see that
\[T\indices{_0^{i'}}=0,\quad T\indices{_i^{0}}=0.\]
For the spatial components, one has
\[T\indices{_a^{a'}}=Y_3(e\indices{_a^{c}})f\indices{_c^{a'}}
+2e\indices{_a^{c}}\epsilon\indices{_3_c^{d}}f\indices{_d^{a'}}.\] The
matrix $(f\indices{_c^{a}})$ is defined here as the inverse of the
matrix $(e\indices{_a^{c}})$.  As soon as we fix the transport of the
frame in time, we can compute the time derivative of the matrix
$(T\indices{_a^{a'}})$; we do this for a particular example
in \Sectionref{sec:backgroundapplication}.  In general, we can expect
that these evolution equations for the matrix $(T\indices{_a^{a'}})$
are non-trivial. In any case, the idea for the ``indirect'' reduction
to \oplo in the Gowdy case is the following: first, substitute the
$Y_3$-derivatives of all tensor field components in the equations by
means of \Eqref{eq:Y3ofsymmetricTensor}. Second, append the evolution
equations for the matrix $(T\indices{_a^{a'}})$ to the system of
equations and, third, evaluate the equations only at $\rho_1=0$. Then,
with respect to the Euler coordinates, all unknowns only depend on $t$
and $\chi$, and the evolution system is closed.  Note, however, that
it depends on the properties of the evolution equations of
$(T\indices{_a^{a'}})$, whether the resulting system of evolution
equations yields a well-posed initial value problem. In the example,
which we discuss later, this is the case.

We remark that under the assumption of \U-symmetry, all results
obtained here also hold for spatial $\SoXSt$-topology. We do not
elaborate on this further; a discussion is given in
\cite{beyer09:Nariai2}.

\subsection{Numerical implementation}
\label{sec:implementation}

\subsubsection{Discretization and our numerical infrastructure}
\label{sec:numinfrastructurenew}
In order to compute approximate solutions of our system of partial
differential equations \Eqref{eq:abstractevolution} by means of a
computer, we need to discretize the equations and the unknowns. Our
analysis before based on Fourier series suggests spectral
discretization \cite{canuto88,boyd} in space with the standard
trigonometric basis. We follow most of the conventions in
\cite{boyd}. In order to keep the presentation as short as possible
here, we do not write down formulas wherever we follow the standard
conventions. In particular, we use the collocation method. For the
spatial grid in any of the spatial dimensions, referred to as $x$, we
set
\begin{equation}
  \label{eq:definegrid}
  x_k=(k+\mu)\frac{2\pi}N,\quad k=0,\ldots,N-1,
\end{equation}
where $N$ is the number of grid points in the chosen spatial
direction. The quantity $\mu\in[0,1[$ is a shift quantity.  For the
standard collocation method which we use, the quantity $N$ must be odd.

For simplicity we use the so-called partial summation algorithm
\cite{boyd} for computing the discrete Fourier transforms (DFT) so
far; however, we plan to switch to the Fast Fourier Transform
algorithm (FFT) \cite{FFT,numericalrecipes}, in order to optimize
performance for high spatial resolutions.

This discretization of the equations and unknowns in space yields a
system of ordinary differential equation (ODE) in time for the
spectral coefficients of the unknowns, or equivalently for the values
of all unknown functions at the spatial grid points. This system of
ODEs is called the semi-discrete system. In order to solve this
numerically, one must discretize time as well. For this, we have
implemented a couple of Runge Kutta (RK) variations described in
\cite{numericalrecipes}; namely, first, the non-adaptive $4$th order
RK scheme, second, the $4$th order ``double-step-adaption'' RK scheme
and, third, the adaptive $5$th order ``embedded'' RK scheme. For those
schemes, the time adaption is always ``global in space'', namely, at a
given time the maximal estimated error at all spatial points is
taken. The parameter $\eta$ is the desired accuracy, according to
Eq.~(16.2.7) in \cite{numericalrecipes}, where it is called
$\Delta_0$.  The lower its value, the stronger is the tendency of the
adaption scheme to decrease the time step $h$. For practical reasons,
we also define a minimal time step $h_{min}$, so that the adaption
scheme is prevented from reaching unpractically small values of $h$.

A sophisticated discussion of errors and convergence in such
discretization approaches is given in \cite{canuto88}. We will not
elaborate on this, in general, very complicated problem
analytically. Instead, we will investigate errors and convergence in
our test applications empirically in \Sectionref{sec:numresults}. The
positive experience, which people have gathered over many years of
research with the collocation method, is summarized in Boyd's
empirical ``assumption of equal errors'' \cite{boyd}, which we decided
to rely on in our numerical work.

It is clear that many classes of problems require adaptive techniques
for the spatial resolution. One particular effect for underresolved
numerical solutions obtained by the collocation method is aliasing. We
have not yet implemented any of the explicit antialiasing recipes,
given for instance in \cite{boyd}. Instead, we use the following
simple global spatial adaption technique so far. At each time step of
the numerical evolution, the program computes the Fourier transform of
one representative unknown function; which one is chosen requires some
experiments. In our applications, where symmetry implies that only one
spatial direction is significant, it is then sufficient to do the
following. From the spectral coefficients of this unknown, the code
determines the ``power'' of the upper third of the frequency spectrum
with respect to the significant spatial direction, divided by the
total power. It is straight forward to generalized this to more
general situations. By ``power'' we mean the sum over the squares of
the modulus of the Fourier coefficients.  We call the result of this
computation the ``adaption norm'' $\normadapt$. Besides adaption
itself, this ``norm'' can also be thought of as a measure for the
aliasing error.  When, during the numerical evolution, $\normadapt$
exceeds a prescribed threshold value, the code stops automatically,
interpolates all quantities to a higher spatial resolution and
continues the run. In each of these adaption steps, we have found it
to be reasonable to almost double the spatial resolution.  In any
case, note that this is a primitive adaption method, since it is
``global in space''. In particular, for solutions, which develop sharp
localized features, a local adaption method in space would be more
desirable. This is a future work project. However, even Gowdy spikes
have been treated with our much simpler method in \cite{beyer:PhD}.

Let us, furthermore, mention the possibility of the Intel Fortran
compiler \cite{Intel} on Intel CPUs, which we have worked on
exclusively up to now, to switch from the standard machine supported
number representation called ``double precision'' with round-off
errors of the order $10^{-16}$ to software emulated ``quad precision''
with round-off errors of the order $10^{-32}$. Indeed, this
possibility is exploited in our applications, as is discussed later.

\subsubsection{Practical issues for spatial
  \texorpdfstring{\Sth}{S3}-topology}
\label{sec:practissuesS3}
In this paper, we discuss two codes for spatial \Sth-topology which
both use the infrastructure above, but with slight differences. For
one of the codes, we assume \U-symmetry with a choice of orthonormal
frame reducing the equations to two spatial dimensions; we call this
code the \tplo-code. For the other code, which we call \oplo-code, we
assume Gowdy symmetry, and suppose that the indirect reduction to one
spatial dimension described before leads to a well-posed initial value
formulation. The particular equations which we implement and study in
both cases are discussed in
\Sectionref{sec:backgroundapplication}. 

In order to apply the results, which we obtained for the properties of
Fourier series of smooth \U-symmetric functions on \Sth
in \Sectionref{sec:analysisfourier} in our discretization approach, we
assume that for any choice of resolution, the solution of the
corresponding discretized equations originates in smooth functions on
\Sth.  Let us suppose, that our evolution equations have the property,
that if the initial data of the continuum problem is smooth, then the
corresponding solution of the continuum equations is smooth. For
instance, this is the case for symmetric hyperbolic systems. Then, if
the initial data is approximated by functions, which originate in
smooth functions on \Sth, and, if all our assumptions about the
coefficients and derivative operators in the equations before hold,
then the solution of the discretized originates in smooth functions on
\Sth for any resolution. This is at least true, as far as we can
neglect errors caused by aliasing and the finite number representation
in our computer. Let us assume just for a moment, that these errors
can be neglected.  In particular, \Eqref{eq:fouriersingular} can be
applied to the \tplo-code directly. For the \oplo-code, we need one
further observation, since the equations are only evaluated at
$\rho_1=0$, and hence there is no information about even and odd
Fourier modes with respect to $\rho_1$. Namely, due to
\Eqref{eq:FNpFourier01}, all coefficients associated with $\cos$-modes
with respect to $\chi$ must correspond to an even mode with respect to
$\rho_1$. Analogously, all coefficients associated with $\sin$-modes
with respect to $\chi$ must correspond to an odd mode with respect to
$\rho_1$. This information is sufficient to use
\Eqsref{eq:fouriersingular} as in the \tplo-case.  Now, for even
$p>0$, there are two ways of computing the coefficients
$\mathfrak{b}_{r,p}^{N}$ and $\mathfrak{c}_{r,p}^N$ of the formally
singular terms. Namely, due to the compatibility conditions
\Eqsref{eq:approximation_comp_cond}, we can write both
\[\mathfrak{b}_{r,p}^{N}=\sum_{n=r}^{N/2}g^N_{2n,p}
=-\frac 12 g^N_{0,p}-\sum_{n=1}^{r-1}g^N_{2n,p},\quad
\mathfrak{c}_{r,p}^N=\sum_{n=r}^{N/2}g^N_{2n-1,p}
=-\sum_{n=1}^{r-1}g^N_{2n-1,p}.\] For odd $p$, there is only one way
of writing these coefficients.  Although each pair is equivalent in
exact computations, there can be a difference numerically.  We refer
to the first way of computing these coefficients as ``up-to-down'',
since we need the information of all high frequencies to compute the
low frequency coefficients recursively. The second variant is called
``down-to-up'', since the information from low frequency coefficients
is used to compute the high frequency coefficients recursively. 

A priori, both ways have the potential to amplify numerical
instabilities.  In particular, although the solution originates in
smooth functions on \Sth at one time of the evolution, this together
with round-off errors and aliasing can cause a drift, so that the form
given by \Eqsref{eq:Fourier_smoothapprox3} are violated eventually. We
have not yet built the special structure of the Fourier series into
our numerical infrastructure. Thus, it is possible, that such errors
accumulate, such that, after some time of evolution, the numerical
solution does not represent a smooth solution on \Sth anymore.
Indeed, we found in our numerical experiments with the \tplo-code in
\cite{beyer:PhD}, that, without precautions, the numerical solution
typically drifts away strongly for both the up-to-down and down-to-up
method. We were not able to pin-point the problem. However, for the
down-to-up method, it turns out to be sufficient, after each time
step, to set all those Fourier coefficients to zero explicitly, which
are supposed to vanish according to \Eqref{eq:FNpFourier01}. With this
manipulation, the numerical evolution becomes stable. In particular,
\Eqref{eq:approximation_comp_cond} stays satisfied within reasonable
error limits, and the code is convergent and able to reproduce
explicitly known solutions. The up-to-down method, however, we were
not able to stabilize.

We can expect that similar practical issues exist for the
\oplo-code. Here, however, we must proceed slightly differently, since
the form given by \Eqsref{eq:Fourier_smoothapprox3} cannot be enforced
explicitly.  In order to control the smoothness of the numerical
solution nevertheless, the idea is to control the unknowns directly at
the coordinate singularities $\chi=0,\pi$ in terms of ``boundary
conditions''\footnote{This is the terminology from
  \cite{garfinkle1999}. We use it despite the fact that there are no
  geometrical boundaries at $\chi=0,\pi$.}. Let the frame $\{e_i\}$ be
$Z_3$-invariant as before, and let $S$ be one of the unknown
$Z_3$-invariant tensor fields. Since $Y_3=\pm Z_3$ on the symmetry
axes, one obtains $Y_3(S_{ij})=0$ there. Exploiting this information
by means of \Eqref{eq:Y3ofsymmetricTensor}, implies a homogeneous
linear algebraic ``boundary system'', which yields the ``boundary
conditions'' for $S$ 
\[
S_{i'j}T\indices{_i^{i'}}+S_{ij'}T\indices{_j^{j'}}=0
\quad\text{at $\chi=0,\pi$}.
\] 
For our particular test case later, we solve the boundary system
in \Sectionref{sec:CFEBoundary}. Let us assume for the moment that we
have solved this system. In general, we would like to have the
possibility of either letting the numerical evolution proceed freely
and just monitoring, how well those boundary conditions are satisfied,
or, if necessary, we would like to enforce the boundary
conditions. The latter means that we set the values of the unknowns at
$\chi=0,\pi$ explicitly to the values implied by the boundary
conditions. In order to make this possible, we modify the spectral
conventions slightly as follows, so that both boundary points $\chi=0$
and $\chi=\pi$ correspond to grid points.  Let $f$ be some unknown
function, which we discretize as
\[f(\chi)=\frac{a_0}{\sqrt{2\pi}}+\frac 1{\sqrt\pi}
\sum_{n=1}^{(M-1)/2}(a_n\cos n\chi+b_n\sin n\chi).\] In the standard
collocation approach, which we use for the \tplo-code, the number of
grid points $N$, according to \Eqref{eq:definegrid}, is odd, and
$M=N$.  Recall that the discrete Fourier transform \cite{boyd} is the
linear map from the values of $f$ at the grid points
$(f(x_0),\ldots,f(x_{N-1}))$ to the Fourier coefficients
$(a_0,b_1,a_1,\ldots,b_{(M-1)/2},a_{(M-1)/2})$, which is bijective for
these choices of $N$ and $M$. This is true for any choice of shift
$\mu$.  For the \oplo-code now, we choose $\mu=0$, any even number
$N$, and set $M=N+1$. In this case $x_0=0$ and $x_{N/2}=\pi$.  One
finds easily that for this, the standard discrete Fourier transform is
the map
\[(f(x_0),\ldots,f(x_{N-1}))\mapsto
(a_0,b_1,a_1,\ldots,b_{N/2-1},a_{N/2-1},0,2a_{N/2}).\] Hence, the map
has the standard properties except for the highest frequency.  The
fact, that the discrete Fourier transform always yields zero for the
highest $\sin$-mode can be understood easily, because the value of
$\sin n\chi$ is always zero for $n=(M-1)/2$ at
$\chi=k\frac{2\pi}N$. The main point is now that this map is
nevertheless invertible.  For spectral differentiation, we ignore the
frequency $n=(M+1)/2$ completely.  In practice, it is expected that
this is not problematic, since the highest frequencies are typically
insignificant. Now, in our numerical experiments, which have so far
restricted to the down-to-up method, we find that the \oplo-code with
this spectral infrastructure is very stable and convergent. This is
true even without enforcing the boundary conditions at all and hence
without any explicit control of the smoothness in regimes where the
solution is relatively smooth. Recall that this is not so for the
\tplo-code. Furthermore, the violations in the boundary conditions
typically converge to zero with increasing resolution.  However, if
the simulation approaches a non-smooth regime of the solution, it
seems often necessary to enforce the boundary conditions; this is
discussed for our test applications.


\section{Analysis of test applications}
\label{sec:testsresults}

Before we test and analyze our numerical method
in \Sectionref{sec:numresults}, we briefly introduce some background
for our test application
in \Sectionref{sec:backgroundapplication}. More details can be found
in our similar discussion in \cite{beyer08:TaubNUT}, where we
emphasize the physical and mathematical ideas and interpret the
results.

\subsection{Background of the test application}
\label{sec:backgroundapplication}

\subsubsection{Physical and mathematical background}
Our aim is to compute cosmological solutions of Einstein's theory of
relativity; in particular we are interested in the strong cosmic
censorship conjecture in Gowdy vacuum solutions of Einstein's field
equations for spatial \Sth- or \SoXSt-topologies
\cite{Gowdy73,Isenberg89,Chrusciel90,garfinkle1999,beyer08:TaubNUT,Stahl02}.
All our discussions assume vacuum and a cosmological constant
$\lambda$, so that Einstein's field equations (EFE) in geometric units
$c=1$, $G=1/(8\pi)$ read
\begin{equation}
  \label{eq:EFE}
  \tilde R=\lambda \tilde g.
\end{equation}
Here $\tilde g_{\mu\nu}$ is the spacetime metric, which is the
fundamental unknown encoding the information about the gravitational
field. Its Ricci tensor $\tilde R$ \cite{hawking} is a $2$nd order
quasi-linear expression in the metric. We will always assume four
spacetime dimensions, that the signature of the metric is Lorentzian
$(-,+,+,+)$, and that Cauchy surfaces, i.e.\ the ``surfaces of
constant time'', are diffeomorphic to \Sth. Furthermore, we suppose
$\lambda>0$. 

In \cite{penrose1963,penrose1979}, Penrose introduced his notion of
conformal compactifications. The idea is to rescale the physical
metric $\tilde g$ by means of a conformal factor $\Omega$, which is a
smooth strictly positive function on the spacetime manifold $\tilde
M$. This yields the so called conformal metric
\[g:=\Omega^2\tilde g.\] Now, loosely speaking, if it is possible to
attach those points to $\tilde M$, which are the limit points of
vanishing $\Omega$, so that the new manifold $M$ is smooth and the
metric $g$ can be extended as a smooth metric on $M$, then we say that
the original spacetime has a smooth conformal compactification. The
references above, but in particular \cite{Friedrich2002}, give further
necessary technical requirements to make this loose statement
rigorous. Under those conditions, the set $\Omega=0$ is a smooth
surface in $M$, called conformal boundary $\scri$. Physically it
represent ``infinity''. In \cite{Friedrich2002}, it is shown, that
conformal boundaries must be spacelike hypersurfaces with respect to
the conformal metric for all solutions of \Eqref{eq:EFE} with
$\lambda>0$.  One calls such solutions ``future asymptotically
de-Sitter'' (FAdS) \cite{DeSitter,galloway2002}, if its conformal
boundary has a smooth non-empty future component $\scrip$; there is
the analogous concept for the past time direction. In particular, the
de-Sitter solution \cite{hawking} is FAdS. Under these conditions,
$\scrip$ represents the infinite timelike future of $\tilde M$. Some
of the asymptotic geometric properties of FAdS solutions are discussed
in \cite{beyer:PhD}.

Friedrich introduced his \term{conformal field equations} (CFE), as
reviewed for instance in \cite{Friedrich2002}, in order to deal with
conformally compactified solutions of Einstein's field equations. In
these conformally invariant equations, the fundamental unknown is the
conformal metric $g$ and the conformal factor $\Omega$ related to the
physical metric $\tilde g$. The non-trivial property of these
equations is, that they are, first, equivalent to Einstein's field
equations wherever $\Omega>0$, and, second, yield regular hyperbolic
evolution equations even where $\Omega=0$.  Under the assumptions
above, the CFE allow us to formulate what we call ``$\scrip$-Cauchy
problem'' \cite{DeSitter}. The idea is to prescribe data for the CFE
on the hypersurface $\scrip$, including its manifold structure,
subject to certain constraints implied by the CFE. These data can then
be integrated into the past by means of evolution equations implied by
the CFE. Friedrich proved that the \scrip-Cauchy problem is
well-posed, and that the unique FAdS solution corresponding to a given
choice of smooth data on \scrip is smooth, as long as it can be
extended into the past.  It is remarkable that this Cauchy problem
allows to control the future asymptotics of the solutions explicitly
by the choice of the data on \scrip. Concerning the past behavior of
the solution corresponding to a given choice of data on \scrip,
however, there is only limited understanding and a-priori control,
because of the complexity of the field equations.  In this paper, we
will give no details on the constraints on \scrip, and say only
briefly what the relevant initial data components are, since we do not
want to introduce all necessary geometric concepts now. However, we
write down a special class of solutions of the constraints
in \Sectionref{sec:id2}. We refer to \cite{DeSitter,beyer:PhD}, where
the details have been carried out.

We decided to use the so-called \term{general conformal field
  equations}, which are the CFE in conformal Gauss gauge
\cite{AntiDeSitter,Friedrich2002}.  In our applications, we specialize
the gauge even further to what we call \term{Levi-Civita conformal
  Gauss gauge} \cite{beyer:PhD}. In this paper here, we will discuss
neither the physical properties, nor the possibly bad implications of
this choice of gauge, but just refer to
\cite{beyer:PhD,beyer08:TaubNUT}. In any case, assuming, without loss
of generality, $\lambda=3$, and having fixed the residual gauge
initial data, as described in \cite{beyer:PhD}, the implied set of
evolution equations is
\begin{subequations}%
  \label{eq:gcfe_levi_cevita_evolution}%
  \begin{align}
    \label{eq:evolution_frame}
    \partial_t e\indices{_a^c}&=-\chi\indices{_a^b}e\indices{_b^c},\\
    \partial_t\chi_{ab}
    &=-\chi\indices{_{a}^c}\chi_{cb}
    -\Omega E_{ab}
    +L\indices{_{{a}}_{b}},\\
    \partial_t\Connection abc
    &=-\chi\indices{_a^d}\Connection dbc
    +\Omega B_{ad}\epsilon\indices{^b_c^d},\\
    \partial_t L_{ab}
    &=-\partial_t\Omega\, E_{ab}-\chi\indices{_a^c}L_{cb},\\
    \label{eq:Bianchi1}
    \partial_t E_{fe}-D_{e_c}B_{a(f}\epsilon\indices{^a^c_{e)}}
    &=-2\chi\indices{_c^c}E_{fe}
    +3\chi\indices{_{(e}^c}E_{f)c}
    -\chi\indices{_c^b}E\indices{_b^c}g_{ef},\\
    \label{eq:Bianchi2}
    \partial_t B_{fe}+D_{e_c}E_{a(f}\epsilon\indices{^a^c_{e)}}
    &=-2\chi\indices{_c^c}B_{fe}
    +3\chi\indices{_{(e}^c}B_{f)c}
    -\chi\indices{_c^b}B\indices{_b^c}g_{ef},\\
    \label{eq:conffactor}
    \Omega(t)&=\frac 12\, t\, (2-t),
  \end{align}
  for the unknowns 
  \begin{equation}
    u=\left(e\indices{_a^b}, \chi_{ab}, \Connection abc, L_{ab}, E_{fe},
      B_{fe}\right).
  \end{equation}
\end{subequations}
The unknowns are the spatial components $e\indices{_a^b}$ of a smooth
frame field $\{e_i\}$ as in \Eqref{eq:orth_frame_Ya}, with
$e_0=\partial_t$, which is orthonormal with respect to the conformal
metric, the spatial frame components of the second fundamental form
$\chi_{ab}$ of the $t=const$-hypersurfaces with respect to $e_0$, the
spatial connection coefficients $\Connection abc$, given by
$\Connection abc e_b=\nabla_{e_a}e_c-\chi_{ac}e_0$ where $\nabla$ is
the Levi-Civita covariant derivative operator of the conformal metric,
the spatial frame components of the Schouton tensor $L_{ab}$, which is
related to the Ricci tensor of the conformal metric by
\begin{equation*}
L_{ij}=R_{ij}/2
  -g_{ij}g^{kl}R_{kl}/12,
\end{equation*}
and the spatial frame components of the electric and magnetic parts of
the rescaled conformal Weyl tensor $E_{ab}$ and $B_{ab}$
\cite{Friedrich2002,FriedrichNagy}, defined with respect to $e_0$.  In
this special conformal Gauss gauge, the timelike frame field $e_0$ is
hypersurface orthogonal, i.e.\ $(\chi_{ab})$ is a symmetric matrix. In
order to avoid confusions, we point out that, in principle, the
conformal factor $\Omega$ is part of the unknowns in Friedrich's
formulation of the CFE. However, for vacuum with arbitrary $\lambda$,
it is possible to integrate its evolution equation in any conformal
Gauss gauge explicitly \cite{AntiDeSitter}, so that $\Omega$ takes the
explicit form \Eqref{eq:conffactor} in our gauge.  We note,
furthermore, that, since $(E_{ab})$ and $(B_{ab})$ are tracefree by
definition, we can get rid of one of the components for each of the
two. Our simple minded choice is the $33$-component by
$E_{33}=-E_{11}-E_{22}$; the same for the magnetic part.  The
evolution equations \Eqsref{eq:Bianchi1} and \eqref{eq:Bianchi2} of
$E_{ab}$ and $B_{ab}$ are derived from the Bianchi system
\cite{Friedrich2002}. In our gauge, the constraint equations implied
by the Bianchi system take the form
\begin{equation}
  \label{eq:bianchi_constraints}
  D_{e_c} E\indices{^c_e}
  -\epsilon\indices{^a^b_e}B_{da}\chi\indices{_b^d}=0,\quad
  D_{e_c} B\indices{^c_e}
  +\epsilon\indices{^a^b_e}E_{da}\chi\indices{_b^d}=0.
\end{equation}
Here, $\epsilon\indices{_a_b_c}$ is the totally antisymmetric symbol
with $\epsilon\indices{_1_2_3}=1$, and indices are shifted by means of
the conformal metric.  The other constraints of the full system above
are equally important, but are ignored for the presentation here.
Further discussions of that evolution system and the quantities
involved can be found in the references above.  Note that in
\Eqsref{eq:Bianchi1}, \eqref{eq:Bianchi2} and
\eqref{eq:bianchi_constraints}, the fields $\{e_a\}$ are henceforth
considered as differential operators, using \Eqref{eq:orth_frame_Ya}
and writing the fields $\{Y_a\}$ as differential operators in the
Euler coordinate basis as in
\Eqsref{eq:coordinate_repr_standard_frameY}. Seen as a system of
partial differential equations, the system
\Eqsref{eq:gcfe_levi_cevita_evolution} is symmetric hyperbolic, and
hence the initial value problem is well-posed.

Note that in this gauge, our initial hypersurface \scrip corresponds
to $t=0$. The past conformal boundary, if it exists, corresponds to
$t=2$. Hence, our time coordinate runs backwards with respect to
physical time.

These equations hold without any symmetry assumptions. In the
following we will assume that all unknown fields involved are Gowdy
symmetric. For the \oplo-code, we need to derive the evolution
equations of the matrix $(T\indices{_a^{a'}})$. The fact that the
conformal metric $g$ is $Y_3$-invariant, implies, after straight
forward computations, that the matrix
$(T_{ab}):=(T\indices{_a^{a'}}g_{a'b})$ is antisymmetric. Our choice
of frame transport is parallel transport with respect to the conformal
metric. This implies, after some algebra, that
\[\partial_t(T\indices{_a^{a'}})=0.\]
Hence, since the original system of equations is symmetric hyperbolic,
also the ``indirect reduction'' to \oplo-dimensions is symmetric
hyperbolic. So, the initial value problem for these equations is also
well-posed.

\subsubsection{A class of initial data}
\label{sec:id2}
As initial data on \scrip, we use the ``Berger data'', which are
solutions of the constraints derived for \U- and Gowdy symmetry in
\cite{beyer:PhD}. Those data are close to data of the
$\lambda$-Taub-NUT solutions and hence are particularly interesting
for the strong cosmic censorship conjecture
\cite{beyer08:TaubNUT}. Here, we restrict to Gowdy symmetry. Under the
conventions above, these data take the form
\begin{subequations}
\label{eq:ID}
\begin{align}
  \label{eq:IDframe}
  (e\indices{_a^b})&=\mathrm{diag}(1,1,a_3),\\
  (\chi_{ab})&=\mathrm{diag}(-1,-1,-1),\\
  \Gamma\indices{_1^1_2}&=0,\quad
  \Gamma\indices{_1^2_3}=-1/a_3,\quad
  \Gamma\indices{_2^1_2}=0, \quad
  \Gamma\indices{_2^1_3}=1/a_3,\\
  \Gamma\indices{_2^2_3}&=0,\quad
  \Gamma\indices{_3^1_2}=1/a_3-2a_3,\quad
  \Gamma\indices{_3^1_3}=0,\quad
  \Gamma\indices{_3^2_3}=0,\\
  (L_{ab})&=\mathrm{diag}\Bigl((5-3/a_3^2)/2,\,\,(5-3/a_3^2)/2,\,\,
  (-3+5/a_3^2)/2\Bigr),\\
  (B_{ab})&=\mathrm{diag}\Bigl(-4(1-a_3^2)/a_3^3,
  \,\,-4(1-a_3^2)/a_3^3,\,\,8(1-a_3^2)/a_3^3\Bigr),\\
  (E_{ab})&=\left(
    \begin{array}{ccc}
      E_0+C_2\,w_{20} & 0 & -\sqrt{2}\,a_3\, C_2\,\Re w_{21}\\
      0 & E_0+C_2\,w_{20} & -\sqrt{2}\,a_3\, C_2\,\Im w_{21}\\
      -\sqrt{2}\,a_3\, C_2\,\Re w_{21} 
      & -\sqrt{2}\,a_3\, C_2\,\Im w_{21} & -2(E_0+C_2\,w_{20})
    \end{array}\right).
\end{align}
\end{subequations}
The induced conformal $3$-metric of \scrip is a Berger sphere with a
free parameter $a_3>0$. The only inhomogeneous, i.e.\ space dependent
part of the initial data is given by the components $E_{ab}$.  For our
definition of the functions $w_{np}$, consult \cite{beyer:PhD}; we
just note that, with respect to the Euler coordinates, we have
\[w_{20}=\cos \chi,\quad w_{21}=\sin \chi\, e^{-i\rho_1}/\sqrt{2}.\]
For all these data, one finds
\begin{equation*}
  (T\indices{_a^{a'}})=
  \begin{pmatrix}
    0 & 2 & 0\\
    -2& 0 & 0\\
    0 & 0 & 0
  \end{pmatrix}.
\end{equation*}
In total this family of solutions of the constraints has three free
parameters $a_3>0$, $E_0\in\R$ and $C_2\in\R$. We remark that the
reason for the strange names of these parameters is consistency with
our notation in \cite{beyer:PhD}.

\subsubsection{Boundary control for the
  \texorpdfstring{\oplo}{1+1}-code}
\label{sec:CFEBoundary}
In \Sectionref{sec:practissuesS3}, we have motivated our boundary
control approach for the \oplo-code. Because the analysis depends
strongly on the particular equations and choice of frame transport, it
was not possible to give a further discussion there in full
generality. Hence, let us continue here for our special choice of
equations and frame transport. Due to what was said before, we have
\begin{equation*}
  (T\indices{_a^{a'}})=
  \begin{pmatrix}
    0 & 2 & 0\\
    -2& 0 & 0\\
    0 & 0 & 0
  \end{pmatrix},
\end{equation*}
for all $t$, for our choice of initial data. In this case, we say,
that the frame is ``boundary adapted''.  Now, we introduce the new
fields
\begin{equation}
  \label{eq:newfields}
  E_{11}^*:=(E_{11}+E_{22})/2, \quad E_{22}^*:=(E_{11}-E_{22})/2,
\end{equation}
and similar for the magnetic part $B_{ab}$, so that the boundary
system, introduced in \Sectionref{sec:practissuesS3}, yields the
following conditions at $\chi=0$ and $\chi=\pi$,
\begin{equation}
  \label{eq:BCs}
  E_{12}=E_{13}=E_{22}^*=E_{23}=0,\quad B_{12}=B_{13}=B_{22}^*=B_{23}=0,
\end{equation}
whereas $E_{11}^*$ and $B_{11}^*$ are free. For all other symmetric
invariant $2$-tensor fields, for instance the $2$nd fundamental form,
we get analogous relations, but, in addition, the 3-3-components are
free, if the tensor is not tracefree. Since the behavior of the
connection coefficients $\Gamma\indices{_a^b_c}$ can be derived on the
symmetry axes as well, which are the only non-tensorial objects in our
set of unknowns, we obtain a complete set of boundary conditions for
all the unknowns. The quantity $\normbc$ is now defined as the sum of
the actual absolute numerical boundary values of all those unknowns,
which are supposed to be zero there according to these
results. Monitoring $\normbc$ in a numerical computation, yields
information on how well the boundary conditions are satisfied.

In order to implement the \oplo-code numerically, we write the
unknowns in terms of the new electric and magnetic fields defined in
\Eqref{eq:newfields}. Actually, it would be better to introduce the
analogous combinations of fields for the other unknowns, but this has
not yet been done. Hence, so far, the code lacks a clean way of
enforcing e.g.\ the boundary condition $\chi_{11}-\chi_{22}=0$ at
$\chi=0$ and $\pi$. To circumvent this problem temporarily, we have
decided to work with a ``partial enforcement'' scheme, which, at a
given time of the evolution, enforces all boundary conditions except
for those of this type. In addition, we monitor the quantity
$\normbc$, and so far this treatment has turned out to be sufficient.

\subsection{Numerical results for the test application}
\label{sec:numresults}

\psfrag{t}[tl][tl][0.7]{$t$}
\psfrag{-ln(0.6952453959-t)}[][][0.7]{$-\ln(0.6952453959-t)$}
\psfrag{diff 1+1 and 2+1}[mc][tc][0.7]{$\normdiff$}
\psfrag{Viol BC}[mc][tc][0.7]{$\normbc$}
\psfrag{norm adapt}[mc][tc][0.7]{$\normadapt$}
\psfrag{Einstein norm}[mc][tc][0.7]{$\normeinstein$}
\psfrag{Bianchi constraint norm}[mc][tc][0.7]{$\normconstr$}
\psfrag{L1(Kretschmann)}[][][0.7]%
{$\left\|\mathrm{Kretschmann}-24\right\|_{L_1}/16$}
\psfrag{Kretschmann}[][][0.7]%
{$\left\|\mathrm{Kretschmann}-24\right\|_{L_1}/16$}

In \cite{beyer:PhD}, we have performed a couple of tests with the
\tplo-code, discussed the findings and drew conclusions about the
numerical method. Here, we rather focus on the \oplo-code, and show so
far unpublished tests and discussions
in \Sectionref{sec:analysis1+1}. Afterwards
in \Sectionref{sec:directcomp}, we also compare a simulation done with
the \tplo-code and the \oplo-code directly.

We just note that we have not made systematic investigations of the
CFL condition for our codes yet.

\subsubsection{Analysis of computations with the
  \texorpdfstring{\oplo}{1+1}-code}
\label{sec:analysis1+1}
\begin{figure}[t]
  \begin{minipage}[t]{0.49\linewidth}
    \centering
    \includegraphics[width=\textwidth]{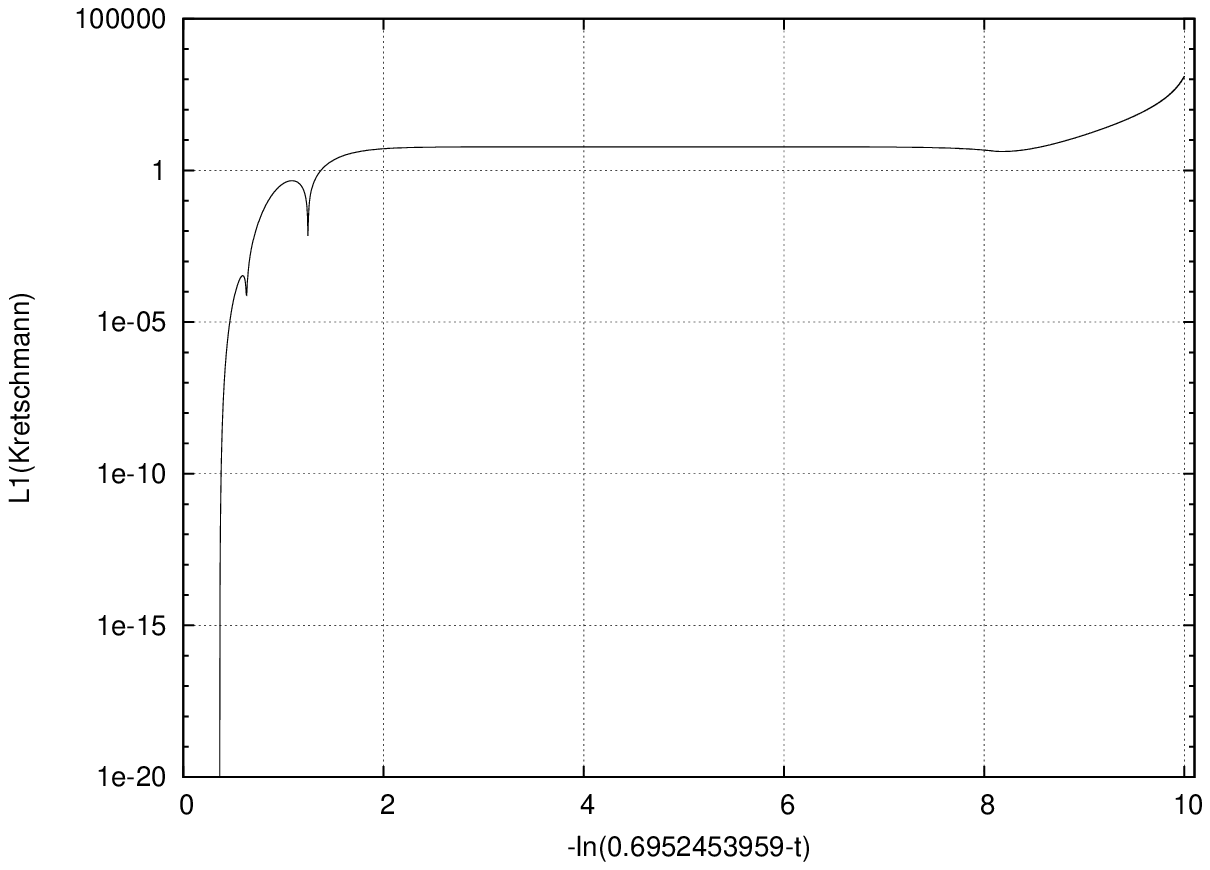}
    \caption{Evolution of curvature for the ``large inhomogeneity case''}
    \label{fig:solution}
  \end{minipage}\hfill
  \begin{minipage}[t]{0.49\linewidth}
    \centering
    \includegraphics[width=\textwidth]{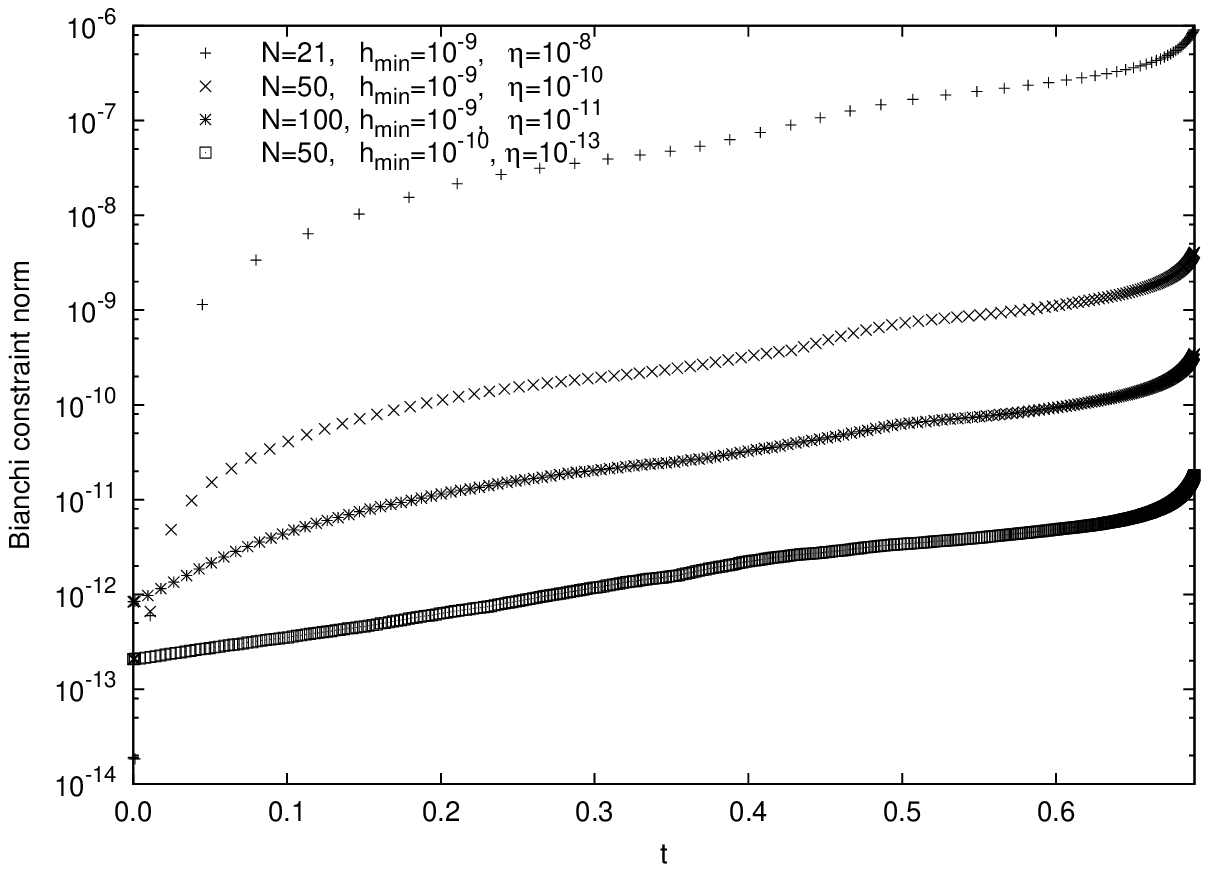}
    \caption{Constraint violations at early times}
    \label{fig:constr_early}
  \end{minipage}
\end{figure}

For our numerical test case here, we choose $a_3=0.7$, $C_2=0.1$ and
$E_0=0$ in \Eqsref{eq:ID} as initial data parameters, corresponding to
the ``large inhomogeneity case'' in \cite{beyer:PhD} and to one of the
simulations presented in \cite{beyer08:TaubNUT}. The associated
solution turns out to develop a singularity, and hence can be seen as
an interesting test case for our code. The evolution of a spatial norm
of the curvature invariant called Kretschmann scalar is shown in
\Figref{fig:solution}.  All the results we show here were done without
the automatic spatial adaption approach described
in \Sectionref{sec:numinfrastructurenew}, because, in order to study
convergence, it seems more useful to control and adapt the spatial
resolution manually. The adaption norm, computed with respect to
$E_{13}$, was used only for estimating the aliasing error. The time
integration was done with the adaptive $5$th order embedded RK scheme
with control parameters $\eta$ and $h_{\min}$ as discussed
in \Sectionref{sec:numinfrastructurenew}. For these runs, we decided
to use the ``partial enforcement'' scheme of the boundary conditions,
explained in \Sectionref{sec:CFEBoundary}. All runs were done with
double precision.

The constraints \Eqref{eq:bianchi_constraints} are satisfied initially
up to machine precision. However, due to numerical errors, those
constraints typically become violated more and more with increasing
evolution time. Let us define $\normconstr$ as the $L^1$-norm of the
sum of the absolute values of each of the six components of the left
hand sides of \Eqsref{eq:bianchi_constraints} at a given instant of
time, all that divided by $\tr(\chi_{ab})$, in order to factor out the
observed collapse of the solution. $L^p$-norms of functions on \Sth
are always evaluated here by means of their corresponding functions on
\T and the standard $L^p$-norm on $\T$. Another norm, which measures
how well the numerical solution satisfies Einstein's field equations,
is
\[\normeinstein
:=\left\|(\tilde R_{ij} -\lambda \tilde
  g_{ij})/\Omega(t)\right\|_{L^{1}(\Sth)},\] where the Ricci tensor
$\tilde R_{ij}$ of the physical metric $\tilde g_{ij}$ is evaluated
algebraically from the conformal Schouton tensor $L_{ij}$ and
derivatives of the conformal factor $\Omega$. The indices in this
expression are defined with respect to the physical orthonormal frame
given by $\tilde e_i=\Omega e_i$. The norm is computed by summing over
the $L^1$-norms of each component.

Now, we will distinguish two phases of the evolution for these initial
data, in which different aspects and effects are important: the early
evolution close to \scrip for $t$ between $0.0$ and $0.69$, and the
late evolution close to the singularity, which we find at
$t\approx0.69520493$. In order to avoid confusions, we recall that the
terms ``early'' and ``late'' are always understood with respect to the
time coordinate $t$, which, however, runs backwards with respect to
the physical time.

\begin{figure}[t]
  \begin{minipage}[t]{0.49\linewidth}
    \centering
    \includegraphics[width=\textwidth]{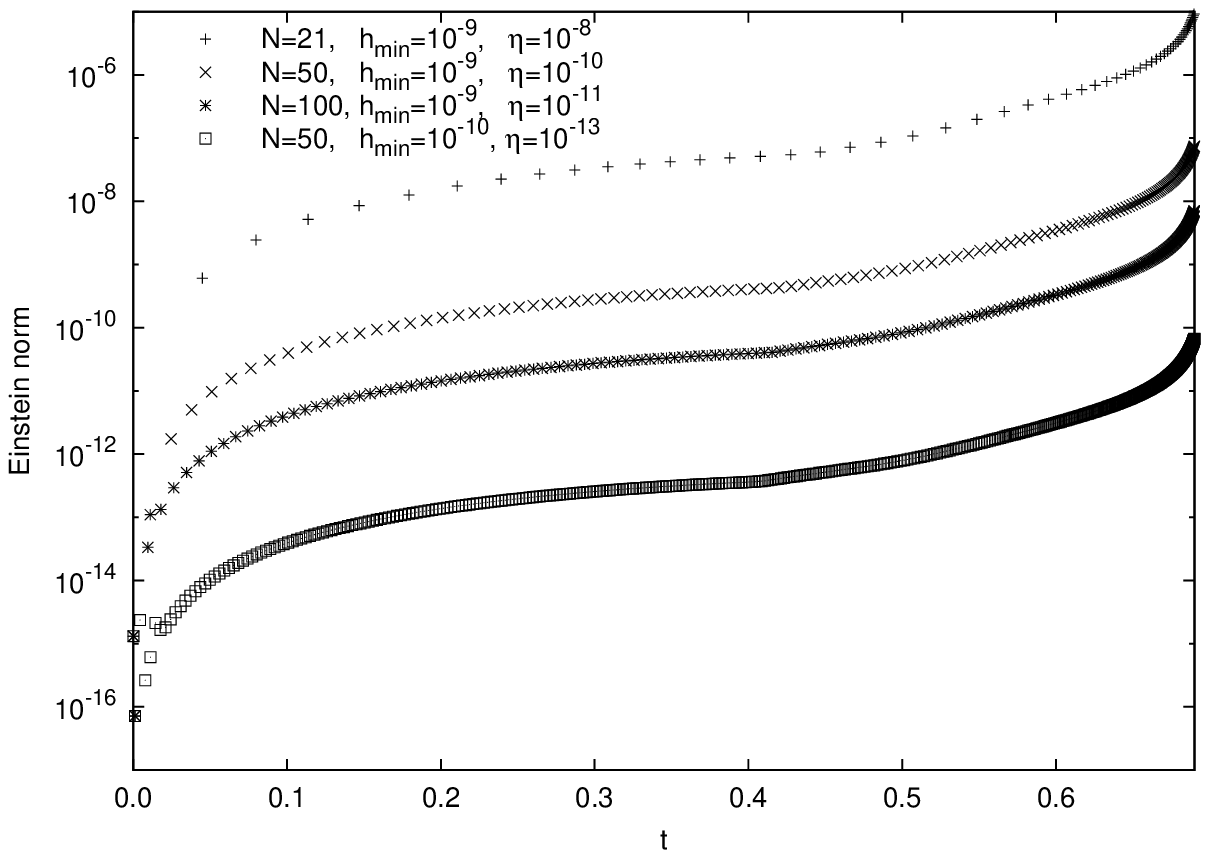}
    \caption{Violations of EFE at early times}
    \label{fig:einstein_early}
  \end{minipage}\hfill
  \begin{minipage}[t]{0.49\linewidth}
    \centering
    \includegraphics[width=\textwidth]{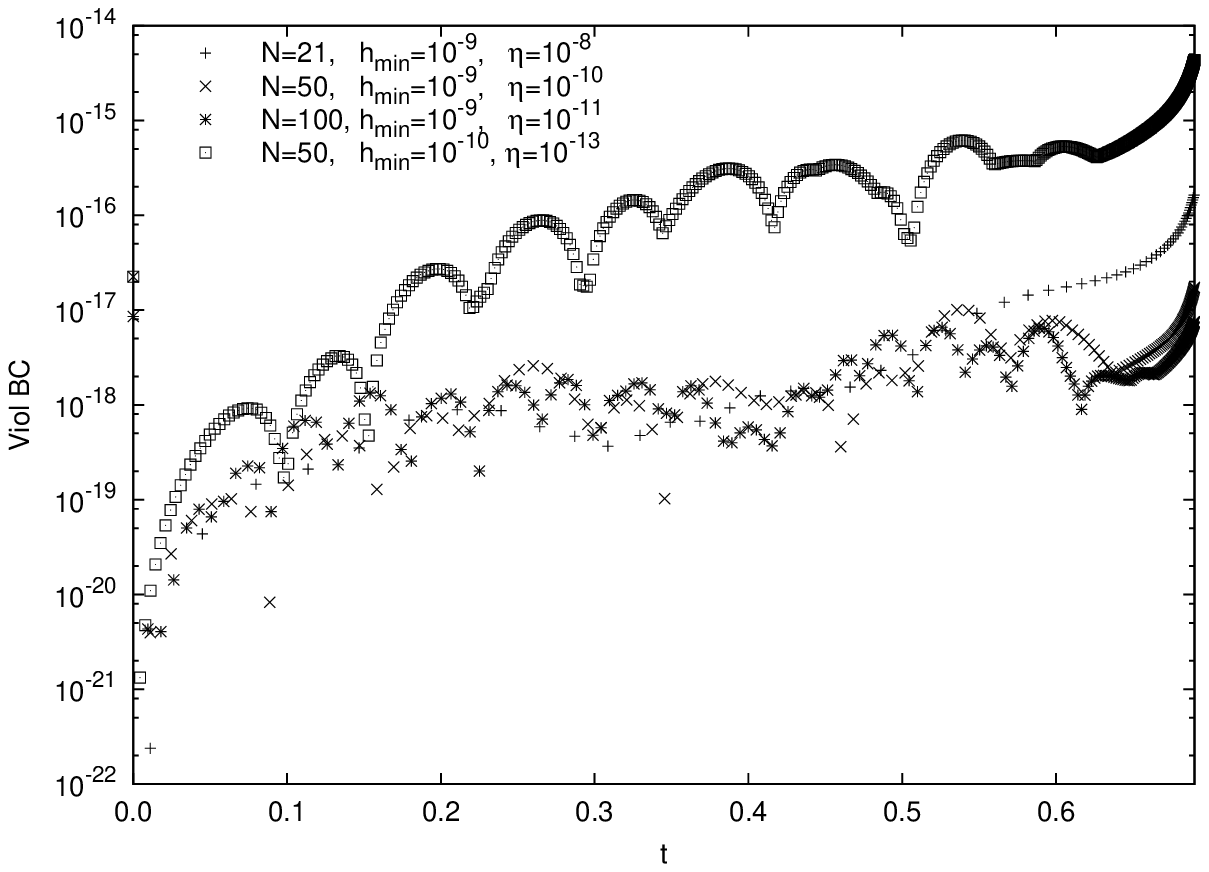}
  \caption{Violation of boundary conditions at early times}
  \label{fig:violbc_early}
  \end{minipage}
\end{figure}

At early times, it is achieved easily that the spatial discretization
error is not significant, until some later time when small spatial
structure starts to form more rapidly. One hint that this is true, as
we do not show here, is that $\normadapt$ is more or less constant
over a long time period, and small. Another hint is that the behaviors
of $\normconstr$ and $\normeinstein$ are not strongly influenced by
the spatial resolution. See \Figref{fig:constr_early} and
\ref{fig:einstein_early}, where $N$ represents the number of spatial
grid points, which is constant in this early regime. Indeed, the
higher $N$, the larger is the initial value $\normconstr$ due to
higher round-off errors for computing spatial derivatives. This is not
visible for $\normeinstein$, since this quantity is defined purely
algebraically in the unknowns.  In \Figref{fig:constr_early}, we see
that $\normconstr$ grows less, the higher the time resolution is,
i.e., in particular, the smaller the parameters $\eta$ and eventually
also $h_{min}$ are. However, we always observe at least weak
approximately exponential growth. In \Figref{fig:einstein_early}, we
see a similar behavior for $\normeinstein$. We do not show here that
there is neither a particular growth of $\normconstr$, nor of
$\normeinstein$, at the symmetry axes. Rather, the maximal growth
takes place, where the curvature increases most strongly. This can be
seen as a confirmation that our treatment of the coordinate
singularities works well, cf.\ \cite{beyer:PhD}. Note that there is an
optimal time resolution, in the sense that, if we choose a higher
resolution, the constraint error and $\normeinstein$ are actually
increased caused by higher round-off errors.  Although we do not show
any plots, we want to mention, that we have experimented with ``quad
precision''. For the \oplo-code, this yields reasonable performance
and has several consequences. First, the initial data for the
constraint violations are decreased by many orders of magnitude, since
those are determined primarily by the precision of the numerical
number representation. By choosing appropriate resolutions, we find
that the constraint violations and $\normeinstein$ can then be kept
several orders of magnitude smaller than in the double precision case
during the whole run. However, they always show exponential growth,
which suggests that this is the typical behavior of the constraint
propagation in our system of equations. Furthermore, quad precision
allows us to work in a regime in which discretization errors are much
larger than round-off errors, and hence it is easier to interpret
convergence tests.
\Figref{fig:violbc_early} shows the behavior of $\normbc$. The errors
at the boundaries are small and stable, despite of some weak growth,
which is expected in situations close to a singularity. In these
tests, this error does not decrease for higher resolutions, which is a
hint that round-off errors have a significant effect. With quad
precision, we were able to confirm that the violations of the boundary
conditions become smaller, consistent with increasing resolution.

\begin{figure}[t]
  \begin{minipage}[t]{0.49\linewidth}
    \centering
    \includegraphics[width=\textwidth]{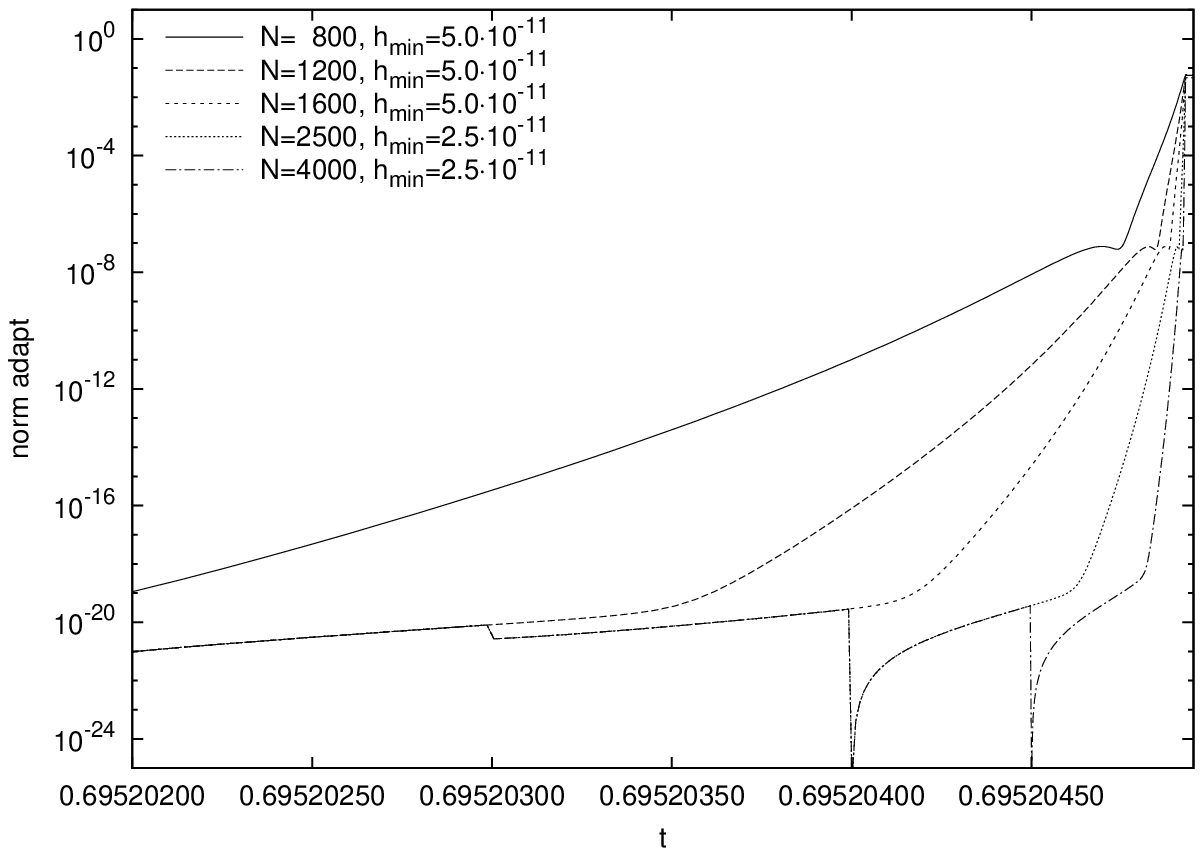}
    \caption{Adaption norm at late times}
    \label{fig:adapt_late}
  \end{minipage}\hfill
  \begin{minipage}[t]{0.49\linewidth}
    \centering
    \includegraphics[width=\textwidth]{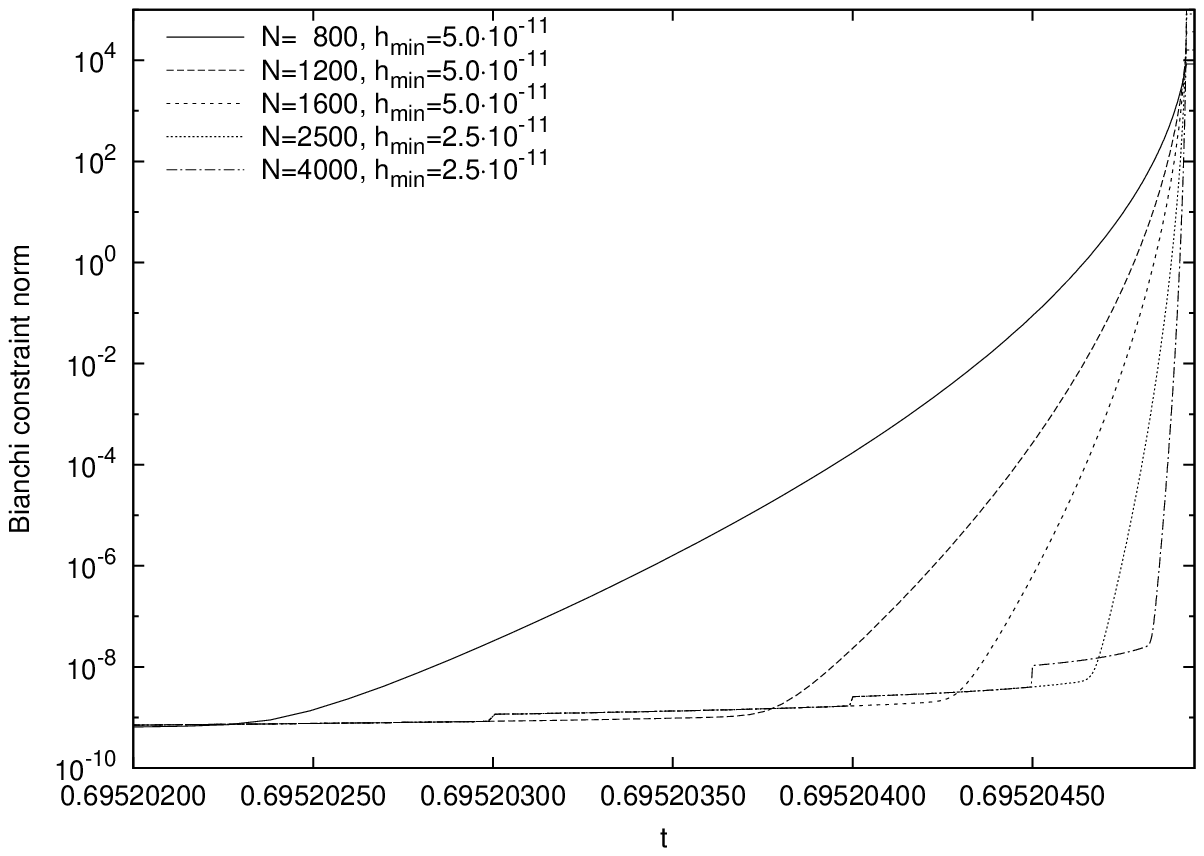}
    \caption{Constraint violations at late times}
    \label{fig:constr_late}
  \end{minipage}
\end{figure}
Concerning the late evolution, the following is found. The following
figures, \Figref{fig:adapt_late} to \Figref{fig:einstein_late}, show a
very small time neighborhood of the final time, where the runs were
stopped and where the solution blows up.  In the runs underlying these
plots, we adapted the spatial resolution several times during the runs
manually; the number $N$ in the figures is the final spatial
resolution in each case. Each manual spatial adaption step is visible
in the plots as a jump, because for different spatial resolutions, the
numerical values of the norms slightly change. In all these runs, we
choose $\eta=10^{-13}$, and hence the time steps $h$ decrease so
strongly that $h=h_{min}$ at that time when the runs were stopped.  In
this late time regime, the errors are dominated by spatial
discretization errors, because the solution has the property that
spatial structures shrink without bound.  This can be seen by looking
at the late time plot of the adaption norm in \Figref{fig:adapt_late}.
It shows, how strongly the demand for spatial resolution grows with
time, but also, that it is possible to gain control by increasing the
resolution at least temporarily. However, the demand for spatial
resolution increases very strongly with time, and it turns into a
difficult numerical issue to keep track of that eventually.  In
\Figref{fig:constr_late}, we demonstrate, how the choice of spatial
resolution influences the propagation of the constraint violations,
and that this quantity converges to a weakly exponentially growing for
sufficiently high spatial resolutions.  This is a promising result,
and shows that the constraint propagation is more or less under
control, as long as it is possible to increase the resolution in
practice.
\begin{figure}[t]
  \begin{minipage}[t]{0.49\linewidth}
    \centering
    \includegraphics[width=\textwidth]{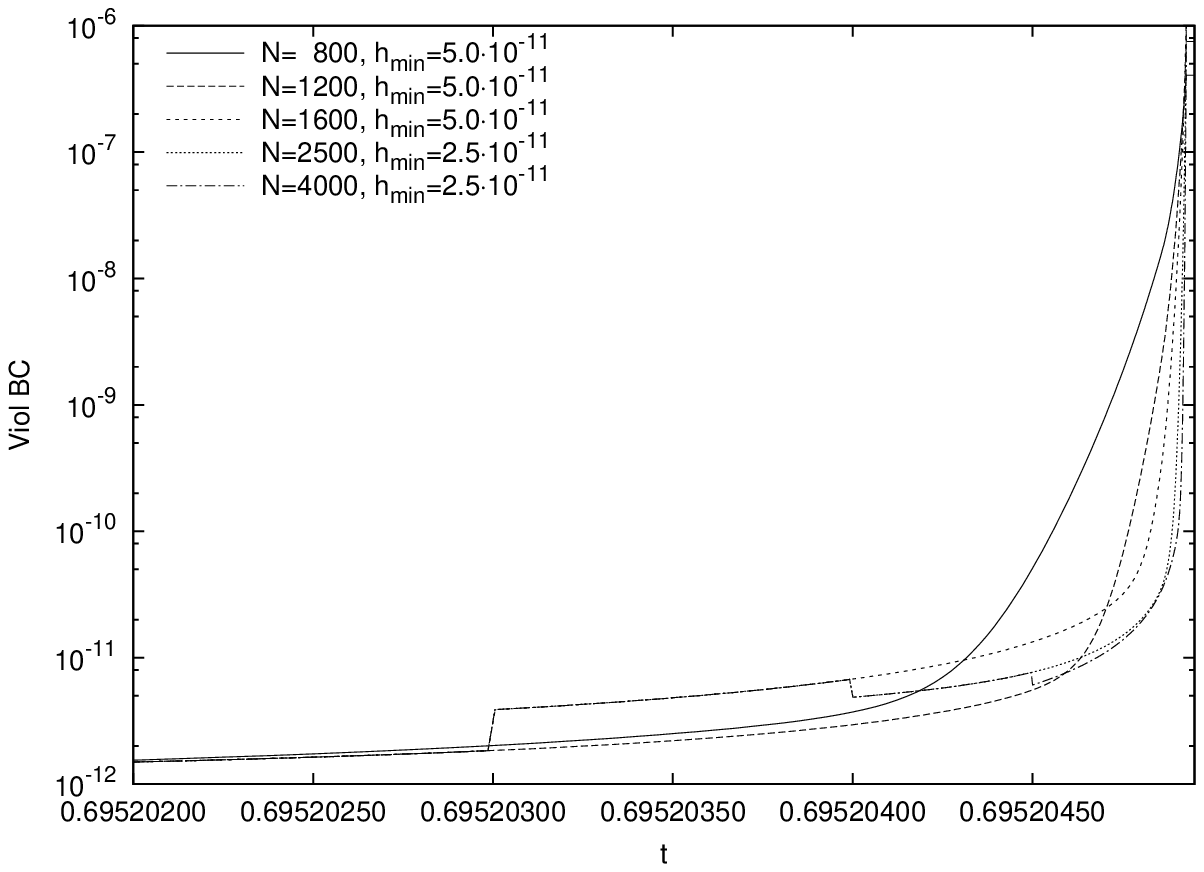}
    \caption{Violation of boundary conditions at late times}
    \label{fig:violbc_late}
  \end{minipage}\hfill
  \begin{minipage}[t]{0.49\linewidth}
    \centering
    \includegraphics[width=\textwidth]{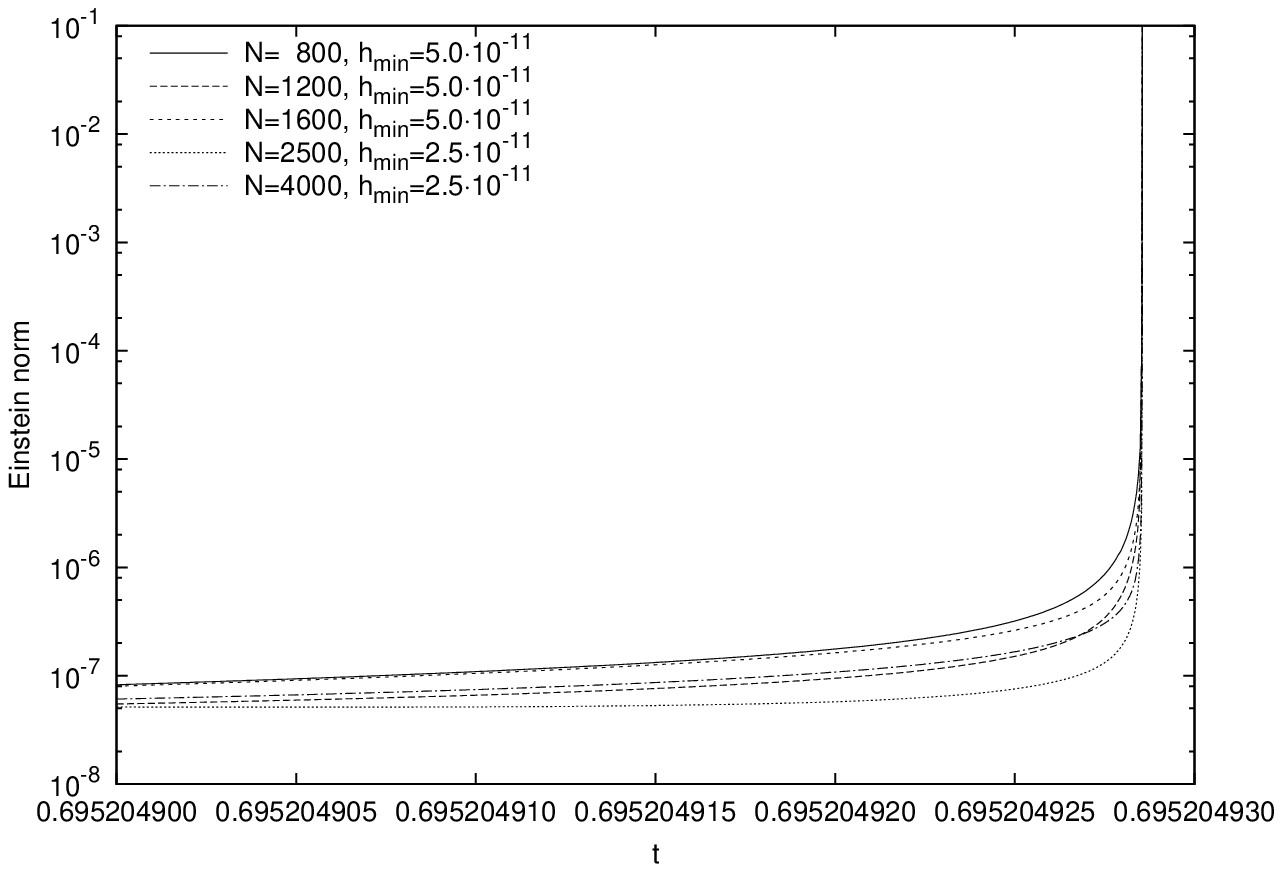}
    \caption{Violations of EFE at late times}
    \label{fig:einstein_late}
  \end{minipage}
\end{figure}
\Figref{fig:violbc_late} shows the violations of the boundary
conditions. They turn out to be very small and under control for
sufficiently large resolutions. The higher the final spatial
resolution is, the smaller these violations appear.  Finally, in
\Figref{fig:einstein_late} we show $\normeinstein$. We do not observe
a very strong difference between the various resolutions; indeed, in
order to make the differences visible at all, the time axis represents
an even smaller time neighborhood now. It is unexpected that at very
late times this norm is not necessarily smaller the higher the
resolution is, and it has to be investigated whether this is a problem.

\subsubsection{Comparison of a computation with the
  \texorpdfstring{\tplo}{2+1}- and \texorpdfstring{\oplo}{1+1}-code
  for a Gowdy symmetric solution}
\label{sec:directcomp}

\begin{figure}[t]
  \centering
  \includegraphics[width=0.49\textwidth]{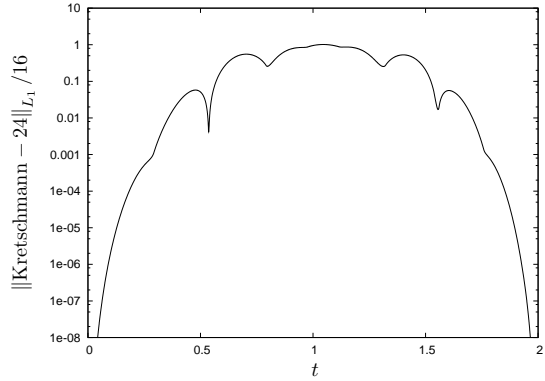}
  \caption{Evolution of curvature for the ``regular case''}
  \label{fig:solution_regular}
\end{figure}

\begin{figure}[t]
  \begin{minipage}[t]{0.48\linewidth}
    \centering
    \includegraphics[width=\textwidth]{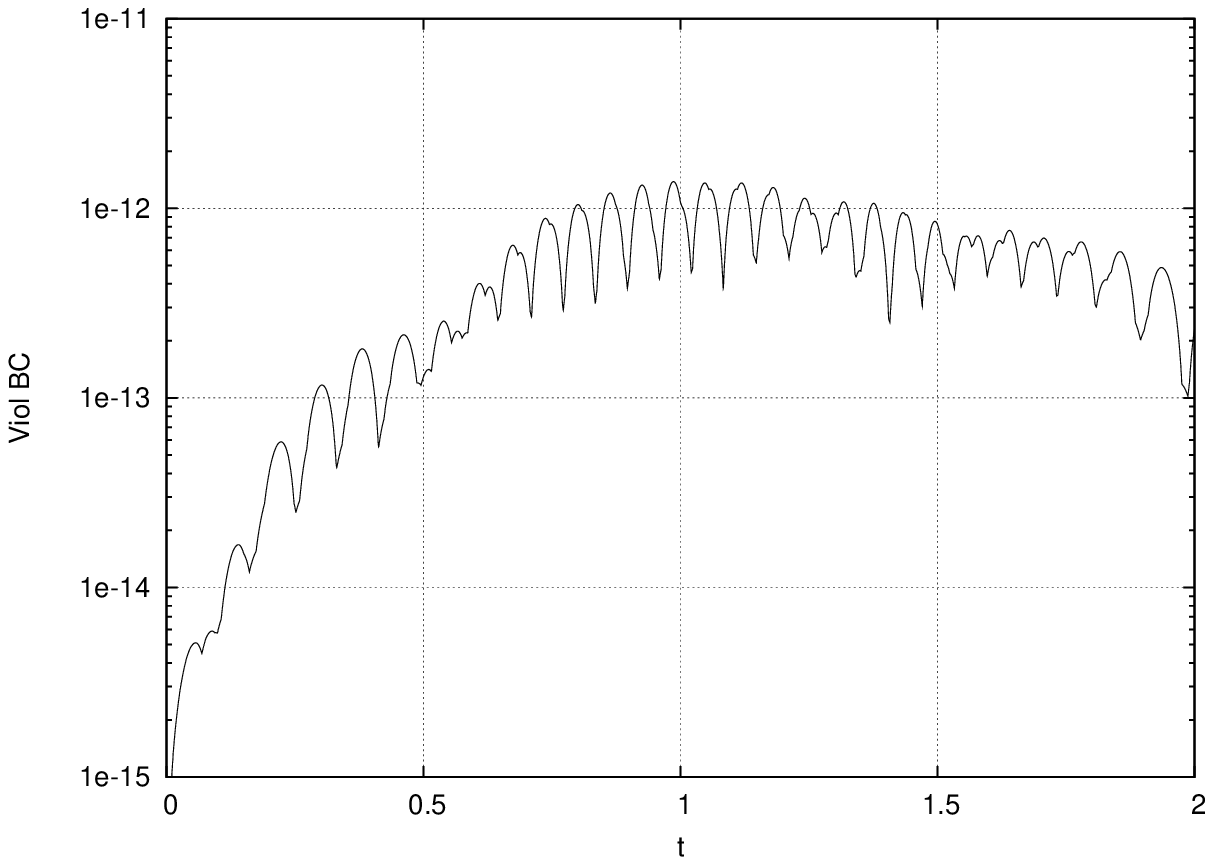}
    \caption{Violations of the boundary conditions of the \oplo-code}
    \label{fig:violBCreg}
  \end{minipage}\hfill
  \begin{minipage}[t]{0.48\linewidth}
    \centering
    \includegraphics[width=\textwidth]{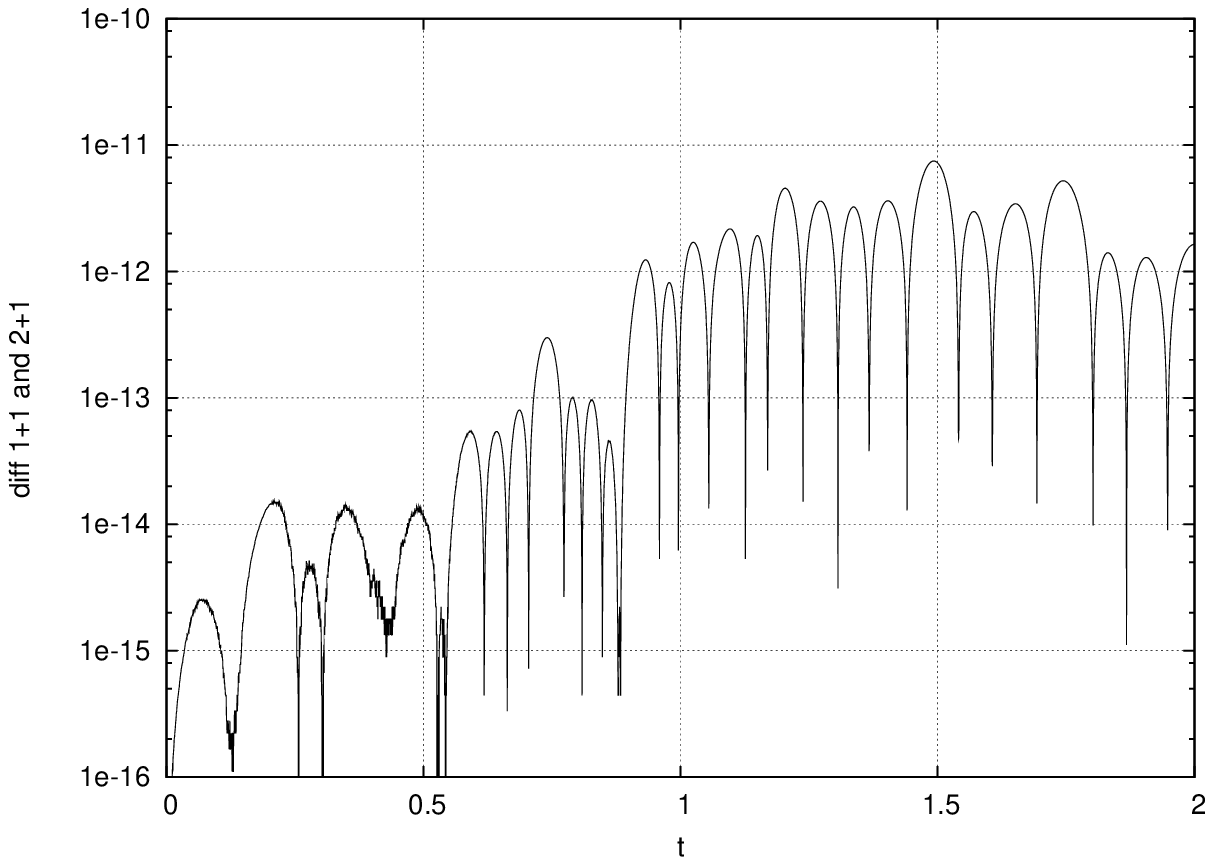}
    \caption{Comparison of the \tplo- and \oplo-code}
    \label{fig:comp2+1_1+1reg}
  \end{minipage}  
\end{figure}
Next, we want to compare directly the numerical results obtained with
the \tplo-code and the \oplo-code for a Gowdy symmetric solution of
our equations. We restrict to a singularity free case, since we do not
want to be spoiled by the lack of spatial resolution, in particular
for the \tplo-code. The following initial data parameters are chosen:
$a_3=0.93$, $E_0=0$ and $C_2=0.5$, which correspond to the ``regular
case'' in \cite{beyer:PhD}. The corresponding solution is smooth for
all $0\le t\le 2$ and hence develops a smooth past conformal boundary
at $t=2$. \Figref{fig:solution_regular} shows the evolution of the
curvature invariant Kretschmann scalar. We show neither convergence
plots, nor the error quantities before, since the situation is very
similar to the early phase of the singular solution in the previous
section. Instead, we report only on one run done with one fixed
resolution: the size of the time step is $h=5\cdot 10^{-4}$ and the
number of spatial points is $N=40$ for the \oplo-code and $N_\chi=41$,
$N_{\rho_1}=21$ for the \tplo-code. We use the non-adaptive $4$th
order RK time integrator, and the automatic spatial adaption has been
switched off here as well.

The run with the \oplo-code was done without enforcing the boundary
conditions now, cf.\ \Figref{fig:violBCreg}. We see that this error
stays very small and is stable. In order to illustrate, how well the
results obtained with the \oplo- and the \tplo-code coincide, let us
define the following norm
\[\normdiff:=\Bigl|(E_{11}^{(\oplo)}|_{\chi=\pi})
-(E_{11}^{(\tplo)}|_{\chi=\pi,\,\rho_1=0})\Bigr|.\] Consider
\Figref{fig:comp2+1_1+1reg} where this norm is plotted vs.\ time.  We
find very good agreement between the two codes. Some deviations can be
expected, since in the \tplo-code, Gowdy symmetry is valid only
approximately; see more comments in \cite{beyer:PhD}. So far, we have
made no effort to explain the oscillatory behavior in both figures,
but we conjecture them to be caused by aliasing, since its amplitude
becomes smaller, the higher the spatial resolution is. We have also
compared more variables at other grid points and found similar
results.


\section{Summary and conclusions}
\label{sec:discussion}
The purpose of this paper is to introduce our numerical approach for
the study of evolution equations with spatial \Sth-topology. First, we
have given details on a geometric point of view in order to find a
natural discretization. Second, we have discussed issues related to
the implementation and, third, analyzed test applications. By all this
we have demonstrated the feasibility of this approach. Indeed, these
techniques have been applied elsewhere, for instance in
\cite{beyer:PhD,beyer08:TaubNUT,beyer09:Nariai2}.

Although our main interest lies in the field of general relativity, we
believe that the applicability of the method is more general.  For the
presentation of this paper, we made a couple of special choices which
need to be overcome in order to use the method in more general
situations. First, our analysis is based on \U-symmetry so far. We
believe that it is straight forward to generalize in
principle. Furthermore, we decided to formulate the equations in terms
of smooth global frames on \Sth. This had the consequence that only
singular terms of special type are present, and with this knowledge,
we were able to work out the Fourier series of the formally singular
terms explicitly in \Sectionref{sec:analysisfourier}. However, in
general, as soon as we have a-priori knowledge about the structure of
the formally singular terms in the equations, and if their type is not
completely different than ours, a similar regularization is
possible. Hence, from this point of view, one should be able to modify
our method to cases, where other kinds of frames or spatial
coordinates are used. In our application, we made a special choice of
coordinate gauge and frame transport. This implied that, in
particular, the matrix $(T\indices{_a^{a'}})$ is constant in time. If
the original evolution equations imply a well-posed initial value
problem, then, in this gauge, also the evolution equations for Gowdy
symmetry reduced to \oplo dimensions have this property.  For other
gauges, however, this might not be the case. A further issue turns up
when the frame transport does not keep the orthonormal frame boundary
adapted in the sense of \Sectionref{sec:CFEBoundary}. Then a different
boundary treatment than ours might become necessary. As a last point
of this certainly incomplete list, we mention that we have not made
experiments with other time integrators yet. Of particular interest
might be implicit schemes, for instance, to treat parabolic evolution
equations.

We believe that our current numerical infrastructure has not yet been
pushed to its limits, but is also not yet really optimized for high
spatial resolutions. For instance, we still do not use FFT but only
the partial summation scheme. Also it may be true, that there is a
more optimal trade-off between accuracy and efficiency for other time
integrators than the Runge Kutta schemes of our choice; comments on
this can be found in \cite{boyd}. Furthermore, it might make sense to
think about parallelization of the code. This should be straight
forward with some publicly available FFT libraries, for instance
\cite{FFTW}. We expect, that the \oplo-code, in its current form, is
limited to a few thousand spatial grid points. We want to stress that
this turned out to be sufficient in our runs with \T-topology in
\cite{beyer:PhD}. There, we were able to reproduce ``spiky features''
numerically. Hence, we are optimistic that also such difficult
phenomena can be studied with our numerical infrastructure in \oplo
dimensions. Dependent on the kind of the application, one may
nevertheless doubt, if spectral methods are suitable at all.  In
general, it is fair to say that, although these methods are highly
accurate for lower resolutions, they might be too inefficient for high
resolutions. Thus it makes sense, to also investigate into other
methods, for instance finite differencing methods, maybe with
multipatch \cite{thornburg04,tiglio05,Lehner05}, in order to make
systematic comparison studies. Further automatic local spatial
adaption, usually called adaptive mesh refinement, would be desirable;
for instance cosmological solutions with spikes have been investigated
with such techniques in \cite{Garfinkle08}, however not for
\Sth-topology.

In summary, due to the results of the tests here and in
\cite{beyer:PhD,beyer08:TaubNUT,beyer09:Nariai2}, we believe that our
numerical approach has promising potentials, which have not been fully
exhausted yet, for many kinds of applications with spatial
\Sth-topology and beyond. For instance, we find it particularly
appealing that our infrastructure can be used directly also for
problems with spatial $\T$-topology and, at least for Gowdy symmetry,
with \SoXSt-topology \cite{beyer09:Nariai2}.  For the applications,
which we have in mind, future research will show, how much the method
must be modified, or whether completely new approaches will become
necessary, in order to deal with probably much more severe phenomena
than those present in the classes of solutions considered so far.


\section*{Acknowledgments}
This work was supported in part by the Göran Gustafsson Foundation,
the Partner Group of the Max Planck Institute for Gravitational
Physics in FaMAF, Córdoba, Argentina, and by the Agence Nationale de
la Recherche (ANR) through the Grant 06-2-134423 entitled Mathematical
Methods in General Relativity (MATH-GR) at the Laboratoire J.-L. Lions
(Universit\'e Pierre et Marie Curie). We would like to thank in
particular Jörg Frauendiener, Helmut Friedrich and David Garfinkle for
discussions and comments, and the relativity group at FaMAF in Córdoba
also for their hospitality.


\bibliography{bibliography}

\begin{thebibliography}{10}

\bibitem{galloway2002}
L.~Andersson and G.J. Galloway.
\newblock d{S}/{CFT} and spacetime topology.
\newblock {\em Adv. Theor. Math. Phys.}, 6:307--327, 2002,
  \urlalt{http://arxiv.org/abs/hep-th/0202161}{hep-th/0202161}.

\bibitem{Bartnik1999}
R.~A. Bartnik and A.~H. Norton.
\newblock {Einstein equations in the null quasi-spherical gauge. {III}:
  {N}umerical algorithms}.
\newblock preprint, 1999,
  \urlalt{http://arxiv.org/abs/gr-qc/9904045}{gr-qc/9904045}.

\bibitem{beyer:PhD}
F.~Beyer.
\newblock {\em Asymptotics and singularities in cosmological models with
  positive cosmological constant}.
\newblock PhD thesis, Max Planck Institute for Gravitational Physics, Sep.
  2007, \urlalt{http://www.arxiv.org/abs/0710.4297}{gr-qc/0710.4297}.

\bibitem{beyer08:TaubNUT}
F.~Beyer.
\newblock Investigations of solutions of {E}instein's field equations close to
  $\lambda$-{T}aub-{NUT}.
\newblock {\em Class. Quant. Grav.}, 25:235005, 2008,
  \urlalt{http://arxiv.org/abs/0804.4224}{arXiv:0804.4224 [gr-qc]}.

\bibitem{beyer09:Nariai2}
F.~Beyer.
\newblock Investigations of the instability of {N}ariai asymptotics within the
  {G}owdy class.
\newblock preprint, 2009,
  \urlalt{http://arxiv.org/abs/0902.2532}{arXiv:0902.2532 [gr-qc]}.

\bibitem{boyd}
J.P. Boyd.
\newblock {\em Chebyshev and Fourier Spectral Methods}.
\newblock Dover Publications, Inc., 2nd edition, 2001.

\bibitem{canuto88}
C.~Canuto, M.Y. Hussaini, A.~Quarteroni, and T.A. Zang.
\newblock {\em Spectral Methods in Fluid Dynamics}.
\newblock Springer, 1988.

\bibitem{Choptuik03}
M.W. Choptuik, E.W. Hirschmann, S.L. Liebling, and F.~Pretorius.
\newblock An axisymmetric gravitational collapse code.
\newblock {\em Class. Quant. Grav.}, 20:1857--1878, 2003,
  \urlalt{http://arxiv.org/abs/gr-qc/0301006}{gr-qc/0301006}.

\bibitem{Chrusciel90}
P.~Chru\'{s}ciel, J.~Isenberg, and V.~Moncrief.
\newblock Strong cosmic censorship in polarized {G}owdy space-times.
\newblock {\em Class. Quant. Grav.}, 7:1671--1680, 1990.

\bibitem{FFT}
J.W. Cooley and J.W. Tukey.
\newblock An algorithm for the machine calculation of complex {F}ourier series.
\newblock {\em Math. Comput.}, 19:297--301, 1965.

\bibitem{tiglio05}
P.~Diener, E.N. Dorband, E.~Schnetter, and M.~Tiglio.
\newblock New, efficient, and accurate high order derivative and dissipation
  operators satisfying summation by parts, and applications in
  three-dimensional multi-block evolutions.
\newblock {\em J. Sci. Comput.}, 32:109--145, 2007,
  \urlalt{http://arxiv.org/abs/gr-qc/0512001}{gr-qc/0512001}.

\bibitem{duez2008}
M.D. Duez, F.~Foucart, L.E. Kidder, H.P. Pfeiffer, M.A. Scheel, and S.A.
  Teukolsky.
\newblock Evolving black hole-neutron star binaries in general relativity using
  pseudospectral and finite difference methods.
\newblock {\em Phys. Rev. D}, 78(10):104015, 2008.

\bibitem{DeSitter}
H.~Friedrich.
\newblock Existence and structure of past asymptotically simple solutions of
  {E}instein's field equations with positive cosmological constant.
\newblock {\em J. Geom. Phys.}, 3(1):101--117, 1986.

\bibitem{AntiDeSitter}
H.~Friedrich.
\newblock Einstein equations and conformal structure: Existence of anti-{D}e
  {S}itter-type spacetimes.
\newblock {\em J. Geom. Phys.}, 17:125--184, 1995.

\bibitem{Friedrich2002}
H.~Friedrich.
\newblock {\em The Conformal Structure of Spacetime: Geometry, Analysis,
  Numerics}, chapter "Conformal Einstein Evolution".
\newblock Lecture Notes in Physics. Springer, 2002.

\bibitem{FriedrichNagy}
H.~Friedrich and G.~Nagy.
\newblock The initial boundary value problem for {E}instein's vacuum field
  equations.
\newblock {\em Commun. Math. Phys.}, 201:619--655, 1999.

\bibitem{FFTW}
M.~Frigo and S.G. Johnson.
\newblock {FFTW} {L}ibrary, \url{http://www.fftw.org/}.

\bibitem{garfinkle1999}
D.~Garfinkle.
\newblock Numerical simulations of {G}owdy spacetimes on {$S^{2}\times
  S^{1}\times R$}.
\newblock {\em Phys. Rev. D}, 60(10):104010, Oct. 1999,
  \urlalt{http://arxiv.org/abs/gr-qc/9906019}{gr-qc/9906019}.

\bibitem{Garfinkle00}
D.~Garfinkle and G.C. Duncan.
\newblock Numerical evolution of {B}rill waves.
\newblock {\em Phys. Rev. D}, 63:044011, 2001,
  \urlalt{http://arxiv.org/abs/gr-qc/0006073}{gr-qc/0006073}.

\bibitem{Garfinkle08}
D.~Garfinkle, W.C. Lim, F.~Pretorius, and P.J. Steinhardt.
\newblock {Evolution to a smooth universe in an ekpyrotic contracting phase
  with $w > 1$}.
\newblock {\em Phys. Rev.}, D78:083537, 2008,
  \urlalt{http://arxiv.org/abs/0808.0542}{arXiv:0808.0542 [gr-qc]}.

\bibitem{Gowdy73}
R.H. Gowdy.
\newblock Vacuum space-times with two parameter spacelike isometry groups and
  compact invariant hypersurfaces: {T}opologies and boundary conditions.
\newblock {\em Ann. Phys.}, 83:203--241, 1974.

\bibitem{hawking}
S.W. Hawking and G.F.R. Ellis.
\newblock {\em The large scale structure of space-time}.
\newblock Cambridge University Press, 1973.

\bibitem{Hennig08}
J.~Hennig and M.~Ansorg.
\newblock {A Fully Pseudospectral Scheme for Solving Singular Hyperbolic
  Equations}.
\newblock preprint, 2008,
  \urlalt{http://arxiv.org/abs/0801.1455}{arXiv:0801.1455 [gr-qc]}.

\bibitem{Intel}
Intel.
\newblock Intel {F}ortran {C}ompiler,
  \urlalt{http://www.intel.com/support/performancetools/fortran/}{http://www.i%
ntel.com/support/performancetools/}.

\bibitem{Isenberg89}
J.~Isenberg and V.~Moncrief.
\newblock Asymptotic behavior of the gravitational field and the nature of
  singularities in {G}owdy space-times.
\newblock {\em Ann. Phys.}, 199:84--122, 1990.

\bibitem{john82}
F.~John.
\newblock {\em Partial Differential equations}.
\newblock Springer, 4th edition, 1982.

\bibitem{Lehner05}
L.~Lehner, O.~Reula, and M.~Tiglio.
\newblock {Multi-block simulations in general relativity: high order
  discretizations, numerical stability, and applications}.
\newblock {\em Class. Quant. Grav.}, 22:5283--5322, 2005,
  \urlalt{http://arxiv.org/abs/gr-qc/0507004}{gr-qc/0507004}.

\bibitem{Majda84}
A.~Majda.
\newblock {\em Compressible Fluid Flow and Systems of Conservation Laws in
  several space dimensions}.
\newblock Springer, 1984.

\bibitem{oneil}
B.~O'Neill.
\newblock {\em Semi-Riemannian Geometry with Applications to Relativity}.
\newblock Academic Press, Inc, 1983.

\bibitem{penrose1963}
R.~Penrose.
\newblock Asymptotic properties of fields and space-time.
\newblock {\em PRL}, 10:66--68, 1963.

\bibitem{penrose1979}
R.~Penrose.
\newblock Singularities and time-asymmetry.
\newblock In S.W. Hawking and W.~Israel, editors, {\em General Relativity -- An
  Einstein Centenary Survey}. Cambridge University Press, 1979.

\bibitem{Pfeiffer02}
H.P. Pfeiffer, L.E. Kidder, M.A. Scheel, and S.A. Teukolsky.
\newblock A multidomain spectral method for solving elliptic equations.
\newblock {\em Comput. Phys. Commun.}, 152:253--273, 2003,
  \urlalt{http://arxiv.org/abs/gr-qc/0202096}{gr-qc/0202096}.

\bibitem{numericalrecipes}
W.H. Press, S.A. Teukolsky, W.T. Vetterlin, and B.P. Flannery.
\newblock {\em Numerical Recipes in C}.
\newblock Cambridge University Press, 2nd edition, 1999.

\bibitem{Rinne05}
O.~Rinne.
\newblock {\em Axisymmetric numerical relativity}.
\newblock PhD thesis, University of Cambridge, 2005,
  \urlalt{http://arxiv.org/abs/gr-qc/0601064}{gr-qc/0601064}.

\bibitem{Scheel06}
M.~A. Scheel, H.~P. Pfeiffer, L.~Lindblom, L.~E. Kidder, O.~Rinne, and S.~A.
  Teukolsky.
\newblock {Solving {E}instein's equations with dual coordinate frames}.
\newblock {\em Phys. Rev. D}, 74:104006, 2006,
  \urlalt{http://arxiv.org/abs/gr-qc/0607056}{gr-qc/0607056}.

\bibitem{Stahl02}
F.~St{\aa}hl.
\newblock Fuchsian analysis of {$S^2\times S^1$ and $S^3$ Gowdy} spacetimes.
\newblock {\em Class. Quant. Grav.}, 19:4483--4504, 2002,
  \urlalt{http://arxiv.org/abs/gr-qc/0109011}{gr-qc/0109011}.

\bibitem{sugiura}
M.~Sugiura.
\newblock {\em Unitary Representations and Harmonic Analysis}.
\newblock North-Holland, Kodansha, 2nd edition, 1990.

\bibitem{thornburg04}
J.~Thornburg.
\newblock Black-hole excision with multiple grid patches.
\newblock {\em Class. Quant. Grav.}, 21:3665--3691, 2004.

\bibitem{Ricciflow}
P.~Topping.
\newblock {\em Lectures on the {R}icci {F}low}.
\newblock Cambridge University Press, 2006.

\bibitem{Wainwright}
J.~Wainwright and G.F.R. Ellis.
\newblock {\em Dynamical Systems in Cosmology}.
\newblock Cambridge University Press, 1997.

\end{thebibliography}
\end{document}